\documentclass[12pt]{article}
\usepackage{epsfig}
\usepackage{amssymb}
\begin{document}

   % title
\begin{center}
   {\Huge Parametrizations of Inclusive Cross Sections for Pion Production 
in Proton-Proton Collisions}\\
   \vspace{.25in}
   {\large Steve R. Blattnig, Sudha R. Swaminathan, Adam T. Kruger, }\\ 
   {\large Moussa Ngom, and John W. Norbury}\\
   \vspace{.15in}
   {\em Physics Department, University of Wisconsin - Milwaukee,}\\
   {\em P.O. Box 413, Milwaukee, Wisconsin 53201, USA.} \end{center}
   \vspace{5mm}

\begin{abstract}
Accurate knowledge of cross sections for pion production in proton-proton
collisions finds wide application in particle physics, astrophysics,
cosmic ray physics and space radiation problems, especially in situations
where an incident proton is transported through some medium, and one
requires knowledge of the output particle spectrum given the input
spectrum. In such cases accurate parametrizations of the cross sections
are desired. In this paper we review much of the experimental data and
compare to a wide variety of different cross section parametrizations. In
so doing, we provide  parametrizations of neutral and charged pion cross
sections which provide a very accurate description of the experimental
data. Lorentz invariant differential cross sections, spectral
distributions and total cross section parametrizations are presented.
  
\end{abstract}

\newpage

\vspace{.15in}
{\bf 1. Introduction}
\vspace{.15in}

Pion production in proton-proton collisions has been extensively studied
over many years, and has now reached the point where this knowledge finds
useful applications in a variety of areas, as detailed below.

1. Two important types of particle detectors are the hadronic and
electromagnetic calorimeters \cite{Fernow}, in which an electromagnetic or
hadronic shower is initiated by a high energy incoming particle. From a
Monte-Carlo simulation of the shower, one is able to deduce important
characteristics of the incoming particle such as its energy and
identity.

2. The primary cosmic rays can be detected by a variety of methods, 
depending on the
incident energy. For the very high energy cosmic rays, where the flux is
relatively low, the extensive air showers (EAS) \cite{Rao, Gaisser, Sokolsky}
provide the most convenient means of detection.  The EAS is analogous to
the hadronic or electromagnetic calorimeter used in particle physics, but
with the Earth's atmosphere being the active volume in which the shower
develops. The EAS has both electromagnetic and hadronic components, and
similar to the calorimeter, the energy and identity of primary cosmic
ray nuclei can be deduced via Monte-Carlo simulation of the showers \cite{Rao,
Gaisser}.

3. In long duration human space flights, such as a mission to Mars, the
radiation levels induced by galactic cosmic rays can exceed exposure
limits set for astronauts \cite{Wilson, Shinn}. In determining the radiation
environment inside a spacecraft  one needs to transport the exterior
cosmic ray spectrum through the spacecraft wall in order to determine the
interior radiation spectrum.

4. In gamma ray \cite{Longair, Dermer} and high energy neutrino astronomy
\cite{Stecker, Protheroe}, the diffuse background radiation is due in large
part to the gamma rays and neutrinos produced in proton collisions with
the protons in the interstellar medium. In addition, pion production from
proton-proton collisions finds applications in the calculation of gamma
ray emission from the accretion disk around a black hole \cite{Mahadevan}.

In all of the above applications it is crucial to have an accurate
knowledge of the cross sections for pion production in proton-proton
collisions. In addition, most of the applications mentioned above require
solving the transport equations which determine the particle spectrum on
one side of a material (active volume of calorimeter, Earth's atmosphere,
spacecraft wall or interstellar medium) given the incident particle
spectrum. Use of pion production cross sections in such transport codes
requires that the cross section be written in a simple form. The transport
codes have many iterative loops, which will take too much computer time if
the cross section formulae also contain many iterative loops. Thus it is
most advantageous if one can write down simple formulae which parameterize
all of the experimental data on pion production cross sections. That is
the aim of the present work.

In this paper we provide simple algebraic parametrizations of charged and
neutral pion production cross sections valid over a range of energies. The
cross sections we provide are Lorentz-invariant differential cross
sections (LIDCS), lab frame spectral distributions (i.e. energy
differential cross sections) and total cross sections, because these are the
types of cross sections most widely used in transport equations. Many
authors have presented such parametrizations before, but the problem is
to decide which authors are correct and whether a particular
parametrization applies only to a limited data set or is valid over a
wider range. In the present work, we have performed an exhaustive data
search and have compared as many different parametrizations as possible
to as much data as possible, so as to provide definitive conclusions as to
which is the most accurate parametrization to use. All of this is
discussed more extensively below.

     The cross sections discussed in this paper are for inclusive pion 
production in proton-proton 
collisions, i.e. the reactions considered are $p+p \rightarrow \pi + X$,
where $p$ represents a proton, $\pi$ represents a pion, and $X$ represents 
any combination of particles. An
extensive search for LIDCS data was performed, and the data was used to
compare all
available parametrizations. An extensive set of data was used in these 
comparisons, but only a few data points are graphed in this paper due to 
space considerations. 
 A method for generating parametrizations
for these cross sections is also described and applied to  $\pi^{\circ}$
production. Spectral distribution and total cross section formulae were
not developed directly because of a lack of data. Instead, the most 
successful LIDCS parametrizations were first
transformed into lab frame spectral distributions by numerical
integration.  These spectral distributions were parameterized and then
numerically integrated to generate lab frame total cross sections.
Finally, the total cross sections were compared to available data and
parameterized as well. This procedure is 
discussed, and
the parametrizations of the numerical results are given.  Multiple checks
of 
the accuracy of all results were made, and some of these are presented.

{\bf Notation:}
Starred quantities (e.g. $\theta^\ast$) refer to the quantities in the center 
of mass (COM) frame, while unstarred quantities (e.g. $\theta$) refer 
to the quantities in the lab frame. 
\newline
$E\frac{d^3 \sigma}{d^3 p} \equiv$ Lorentz-Invariant Differential Cross 
Section  (LIDCS)
\newline
$\frac{d \sigma}{dE} \equiv$ spectral distribution  =  
$2\pi p \int_0^{\theta_{max}} d\theta E\frac{d^{3}\sigma}{d^{3}{p}}\sin\theta$ 
\newline
$\sigma \equiv$ total cross section = $2\pi  \int_0^{\theta_{max}} 
d\theta \int_{p_{min}}^{p_{max}} dp E\frac{d^{3}\sigma}{d^{3}{p}} 
\frac{p^{2}\sin\theta}{\sqrt{p^{2} + m_{\pi}^2}}$ 
= $\int_{E_{min}}^{E_{max}} \frac{d \sigma}{dE} dE$
\newline
$P_p$ is the proton momentum.
\newline   
$m_p$ is the proton mass.
\newline
$m_\pi$ is the pion mass.
\newline
$\sqrt{s}$ is the magnitude of the total four momentum, and is equal to the total energy in the COM frame.
\newline
$T_{lab}$ is the lab frame kinetic energy of the incoming proton.
\newline
$T$ is the pion kinetic energy.
\newline
$E$ is the pion total energy.
\newline
$\theta$ is the angle of pion scattering with respect to the direction of the incident particle.  
\newline
$p$ is the pion momentum.
\newline
$p_\perp \equiv p_t$ is the pion transverse momentum ($p_\perp=p\sin \theta$).
\newline
$p_{max}$ is the maximum possible momentum the scattered pion can have for
a given $\sqrt{s}$.

   \vspace{.15in}
\noindent
{\bf 2. Comparison of Lorentz Invariant Differential Cross Sections}
\vspace{.15in}

The object has been to determine an accurate parameterization for
inclusive LIDCS, which can be confidently applied to regions where no
experimental data is available.  For example, the parametric equation
would need to be extrapolated to energies lower than those for which  data
are available, if the formulae were to be used for the purpose of 
developing radiation shielding materials. The most convenient formulae are
those that are in closed form, since they are easily used, and take
relatively little CPU time in numerical calculations.  Some of the
formulae that were considered as representations of the LIDCS were
not in closed form, but  included tabulated functions of energy (i.e.
numerical values were given for specific energy values rather than a
functional form). When comparing parameterizations, closed form
expressions were given precedence over other equally accurate formulae. 

The invariant single-particle distribution is defined by
\begin{equation}
f(AB \rightarrow CX)\equiv E_c\frac{d^{3}\sigma}{d^{3}{
p}_{c}}\equiv E\frac{d^3\sigma}{d^3p} 
=\frac{E}{p^2}\frac{d^{3}\sigma}{dp d\Omega}
\label{1-1}
\end{equation}
where $\frac{d^{3}\sigma}{d^{3}{p}_{c}}$ is the differential cross-section
(i.e. the probability per unit incident flux) for detecting a particle $C$
within the phase-space volume element $d^{3}{p}_{c}$. $A$ and $B$ are the
initial colliding particles, $C$ is the produced particle of interest, and
$X$ represents all other particles produced in the collision. $E$ is the total
energy of the produced particle $C$, and $\Omega$ is the solid angle. 
 This form is favored since the quantity is invariant under Lorentz
transformations.

The data for pion production in proton-proton interactions is primarily
reported in terms of the kinematic variables $\theta^{\ast}, \sqrt s,
p_\perp$, which are respectively,  the center of mass (COM) frame
scattering angle of the pion, the invariant mass of the entire system, and the transverse momentum of the
produced pion.  $ \sqrt s$ is a Lorentz invariant quantity, and is equal
to the total energy in the COM frame.  $p_\perp \equiv p^\ast
\sin \theta^\ast$, where $p^\ast$ is the COM momentum. $p_\perp$
 is invariant under the transformation from  the lab frame
to the COM frame. 
(See \cite{Byck} for a more detailed discussion of
kinematic variables.) In the following discussions, all momenta,
energies, and masses are in units of GeV.

\vspace{.15in}
\noindent
{\bf 2.1 Neutral Pions}
\vspace{.15in}

\begin{figure}[htb]
\centering
\psfig{file=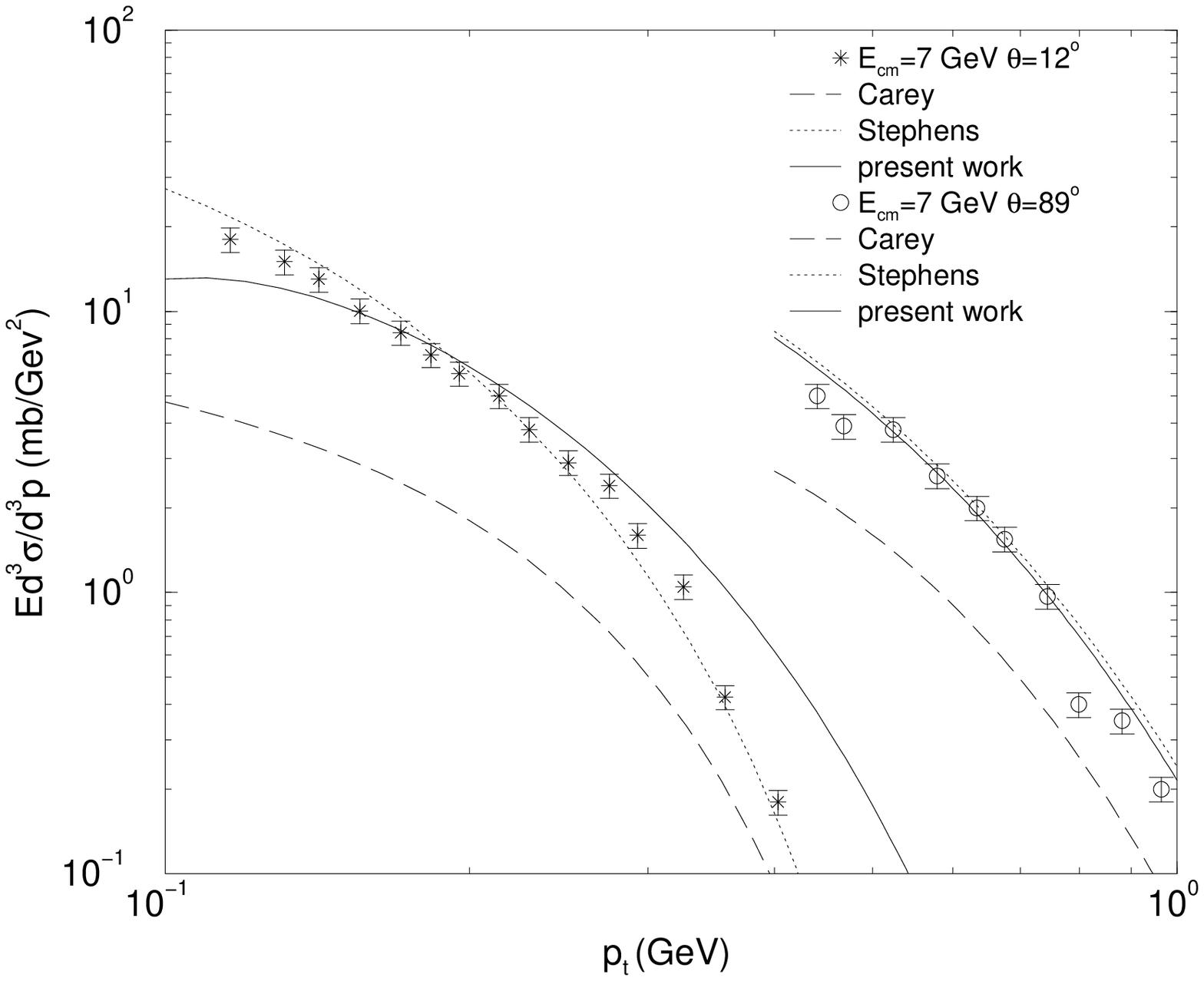,height=3in,width=6in}
\caption{$\pi^0$ production parametrizations of Carey et al. (equation 5) 
\cite{Carey76}
, of Stephens et al. (equation 6) \cite{Stephens}, and of equation (7) plotted with 
LIDCS data from \cite{Fidecaro,Stephens}. LIDCS is plotted against transverse momentum for 
COM energy $E_{cm}=7$ GeV. The pion COM scattering angle is $12.2 < 
\theta^\ast < 12.4^\circ$  and $\theta^\ast=89^\circ$ for the data, and the
parametrizations are plotted 
at $\theta^\ast=12^\circ$ and  $\theta^\ast=89^\circ$.} \label{6.7new}
\end{figure}

\begin{figure}[htb]
\centering
\epsfig{file=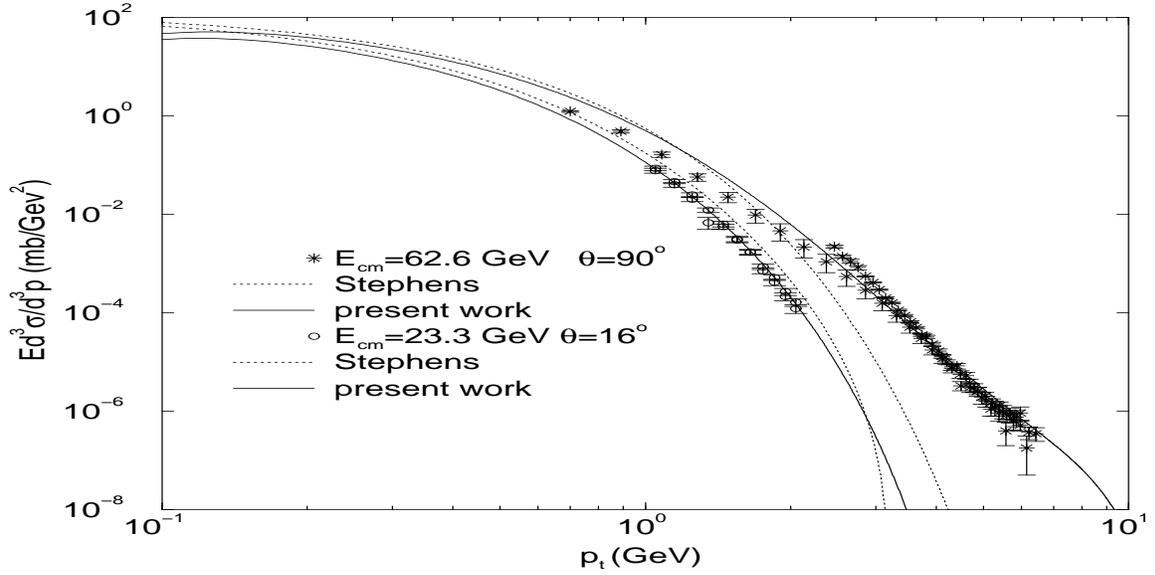,height=3in,width=6in}
\caption{$\pi^0$ production parametrizations of Carey et al. (equation 5) \cite{Carey76}, of Stephens et al. (equation 6)  \cite{Stephens}, and of equation (7) plotted with LIDCS data from \cite{Busser73,Busser76,Eggert}. The data is for COM energy $E_{cm}=62.4$ GeV and 62.9 GeV at  COM scattering angle $\theta^\ast = 90^{\circ}$ 
and the parametrizations are plotted at $E_{cm}=$ 62.6 GeV at  $\theta^\ast = 90^{\circ}$. The second set of data is at COM energy $E_{cm}=23.3$ GeV and the pion COM scattering angle $\theta^\ast = 15^{\circ}$ and $17.5^{\circ}$ for the data, and the parametrizations are plotted at $\theta^\ast=16^\circ$. } \label{62.6new}
\end{figure}

\begin{figure}[htb]
\centering
\epsfig{file=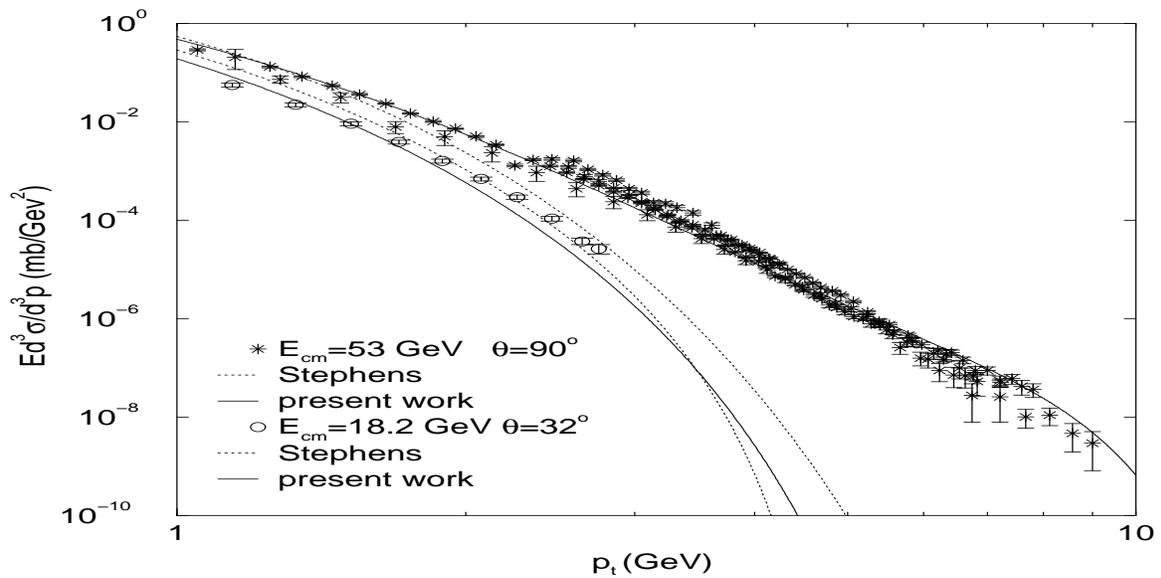,height=3in,width=6in}
\caption{$\pi^0$ production parametrizations of Carey et al. (equation 5) 
\cite{Carey76}, of Stephens et al. (equation 6) \cite{Stephens}, and of equation (7) plotted with LIDCS data from \cite{Carey76}. LIDCS is plotted against transverse momentum for COM energy $E_{cm}=18.2$ GeV. The pion COM scattering angle $32.3^{\circ} < \theta^\ast < 32.5^{\circ}$ for the data, and the parametrizations are plotted at $\theta^\ast=32^\circ$. The second set of data and parametrizations are at $E_{cm}=53$ GeV, $\theta^\ast = 90^{\circ}$. The data is from \cite{Busser73,Busser76,Eggert,Lloydowen}.} \label{53new}
\end{figure}

Busser {\em et al.} \cite{Busser73} have fitted the LIDCS data obtained in the
reaction $p + p \rightarrow \pi^{\circ}+X$, where $p$ represents a proton,
$\pi^{\circ}$ represents the neutral pion produced, and $X$ represents all
other produced particles, to an equation of the form 
\begin{equation}
E\frac{d^3 \sigma}{d^3 p} = Ap_{\perp}^{-n} \exp(-b \frac{p_\perp}{\sqrt{s}})
\label{1-2}
\end{equation}
with $A=1.54 \times 10^{-26}$, $n=8.24$, and $b=26.1$.
This equation is based on a specific set of experimental data with
all measurements taken at  $\theta^\ast \simeq 90^{\circ}$, and was
originally intended only for pions with high $p_\perp$.
Comparison of this parameterization with data available from other
experiments \cite{Fidecaro} - \cite{Stephens} indicates that the global
behavior of the invariant cross section cannot be represented by a 
function of this form. See Figures \ref{6.7new} - \ref{53new} for some
 examples of data.   
The parameterization of Busser {\em et al.} \cite{Busser73} was not plotted
because the cross section is much too 
small compared to the data in the $p_\perp$ ranges covered by the graphs. 

The following form has been used by Albrecht {\em et al.} \cite{Albrecht} 
to represent neutral pion production.
\begin{equation}
E\frac{d^3\sigma}{d^3p}=C(\frac{p_0}{p_\perp + p_0})^n
\end{equation}
where {\em C}, {\em n}, and $p_0$ are free parameters.
This equation only has dependence on $p_\perp$ where as the data 
\cite{Fidecaro}  - \cite{Stephens},
some of which is shown in Figures  \ref{6.7new} - \ref{53new},  also has 
dependence on $\sqrt{s}$ and $\theta^\star$. This form is therefore 
not general enough to represent all the data.

Ellis \cite{Ellis77} have favored a representation for the
invariant cross section of the form 
\begin{equation}
E\frac{d^{3}\sigma}{d^3 p}=A(p_{\perp}^{2}+M^{2})^{-N/2}f(x_{\perp},\theta^{*})
\label{1-3}
\end{equation}

\noindent
where $f(x_{\perp},\theta^\ast)=(1-x_{\perp})^{F}$, $N$ and $F$ are
free parameters and the scaling variable $x_{\perp}$ is given by
$x_{\perp}=\frac{p_{\perp}}{p^*_{max}}  \simeq \frac{2p_{\perp}}{\sqrt{s}}$.
$p^*_{max}=[(s+m_{\pi}^{2}-4m_{p}^2)^{2}/4s-m_{\pi}^{2}]^{\frac{1}{2}}$,
where $m_{\pi}$ and $m_{p}$ are the mass of the neutral pion and the
proton respectively, is the maximum pion momentum allowed. 
The outline of this basic form has been used by Carey
{\em et al.} in fitting the invariant cross section for the inclusive
reaction
$ p + p \rightarrow \pi^{\circ}+X$ \cite{Carey1pi0}. Their
representation
is given by
\begin{equation}
E\frac{d^{3}\sigma}{d^3 p}=A(p_{\perp}^2+0.86)^{-4.5}(1-x^*_R)^{4}
\label{1-4}
\end{equation}
where $x^*_{R}=\frac{p^*}{p^*_{max}}$ is the radial scaling variable and the
normalization constant $A$ has been determined as $A\simeq 5$.
This parameterization accurately reproduces the data  for 
measurements taken at $\theta^\ast =90^{\circ}$ and  $\sqrt{s} \geq 9.8$ 
GeV , but does not
 agree well with the data for lower energies ($\sqrt{s} = 7$ GeV).
The disagreement at lower energies can be seen in Figure \ref{6.7new}.

\begin{figure}[htb]
\centering
\psfig{file=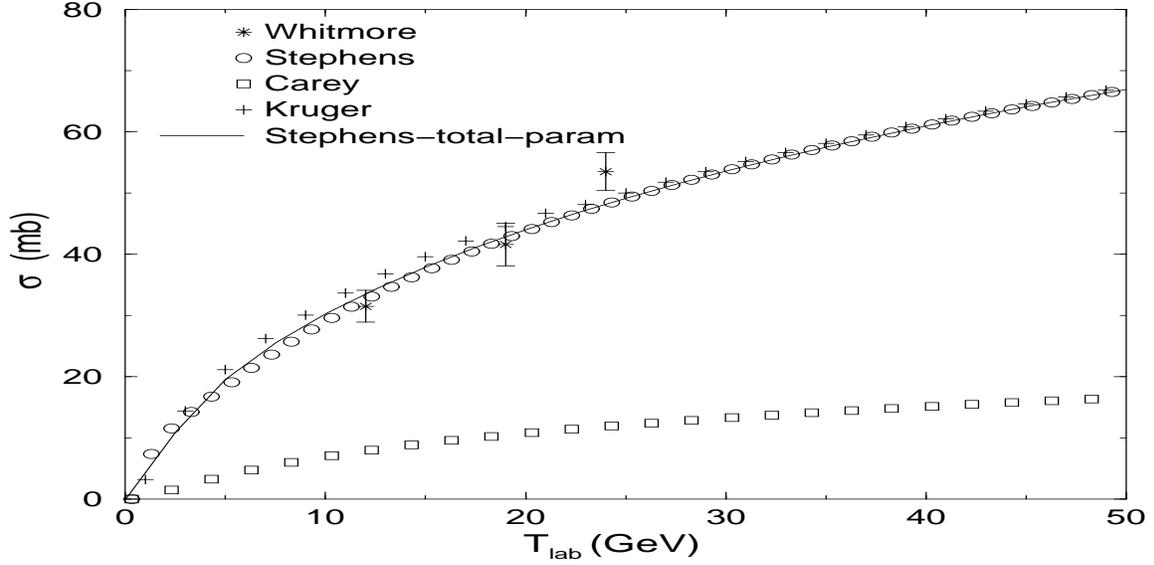,height=3in,width=6in}
\caption{Parametrization of total $\pi^\circ$ production cross section plotted with {\em numerically integrated} LIDCS parametrizations of Stephens et al. (equation 6) \cite{Stephens}, of Carey et al. (equation 5) \cite{Carey76}, and of equation(7) referred to as Kruger. The curve labelled `Stephens-total-param' is the parametrization given in equation (29). Three data points from Whitmore \cite{tcsdat} are included for comparison.}
\label{tcspi0}
\end{figure}

Another problem with this parameterization becomes apparent, when one considers
that integration over all allowed angles and outgoing particle momenta
should yield the total inclusive cross section.  The details of this 
calculation appear in Section 3.  A comparison of the
experimentally determined total cross section data from Whitmore 
\cite{tcsdat} with the results
of the numerical integration of equation [\ref{1-4}] shows that the total cross
section is greatly underestimated by Carey. See Figure \ref{tcspi0}.

Stephens and Badhwar \cite{Stephens} obtained data from the photon cross 
sections given by 
Fidecaro \cite{Fidecaro}. The Fidecaro data was taken at incident 
proton kinetic energy  $T_{lab} = 23$ GeV and
$p_\perp =$ 0.1 GeV - 1.0 GeV. (Note: No error 
was listed by Fidecaro {\em et al.} \cite{Fidecaro} for pion production.  
Error bars of ten percent were added to the data on 
the figures, since this level of error was standard for most of the other 
data. Also, Stephens uses the notation $E_p$
instead of $T_{lab}$.)  Figures  \ref{6.7new} - \ref{53new},
show common examples of the accuracy of Badhwar and 
Stephens' parameterization's fit to the data. 
The following is the parameterization of the $\pi^{\circ}$ invariant cross
section 
proposed by Stephens and Badhwar \cite{Stephens};
\begin{equation}
E\frac{d^3\sigma}{d^3p}
=Af(T_{lab})(1-\tilde{x})^q\exp(-Bp_{\perp}/(1+4m_p^2/s))
\label{1-5}
\end{equation}
\noindent
where
\begin{eqnarray}
\tilde{x} &=& \surd\{(x_{\parallel}^{\ast})^{2}+(\frac{4}{s})(p_{\perp}^2+
m_{\pi}^2) \} \nonumber \\
q &=& \frac{C_1-C_2p_{\perp}+C_3p_{\perp}^{2}}{\sqrt{1+4m_p^2/s}} \nonumber \\
f(T_{lab}) &=& (1+23T_{lab}^{-2.6})(1-4m_p^2/s)^2 \nonumber
\end{eqnarray}
\noindent
and
\begin{center}
$A=140,B=5.43,C_1=6.1,C_2=3.3,C_3=0.6$ with $x_{\parallel}^\ast\equiv
\frac{p_{\parallel}^{\ast}}{p^{\ast}_{max}}$, and $p_{\parallel}^{\ast}=p^{\ast}\cos\theta^{\ast}$.
\end{center}

The Stephens-Badhwar parameterization was found to be the best of the
previously listed representations,
because it accurately reproduces the data in the low $p_{\perp}$ region,
where the cross section is greatest (see Figures \ref{6.7new} - 
\ref{53new}), and its integration yields accurate
values for the total  cross section (see Figure \ref{tcspi0}). This equation 
is, however, a poor tool
for predicting values of the invariant cross section for $p_{\perp} \gtrsim 
3$ GeV, where the value predicted underestimates experimental
data by up to $\simeq10$ orders of magnitude (see Figures \ref{62.6new} 
and \ref{53new}).

No parameterization currently exists that accurately fits the global
behavior of the LIDCS data.  Previous equations have suffered from being
too specific to a particular set of experimental data, or from failing to
reproduce the total cross section upon integration.  It is for these
reasons that a new parameterization is desired, one that correctly
predicts all available data while maintaining the essential quality of
correctly producing the total cross section upon integration.

The approach that has been adopted in the present work is to assume the
following form for the invariant cross section
\begin{equation}
E\frac{d^3\sigma}{d^3p}=(\sin\theta^\ast)^{D(\sqrt s, p_{\perp},
\theta^\ast)}F(\sqrt s, p_{\perp},\theta^\ast = 90^{\circ})
\label{1-6}
\end{equation}
The motivation for an equation of this form is that as the angle decreases,
the cross
section decreases very slowly at lower $p_{\perp}$ values.  The
approximation that was made in deriving the above equation is that as
$p_{\perp}\rightarrow 0$, the cross section is assumed to be independent
of angle.

Under the assumption that the invariant cross section can be
fitted by equation (\ref{1-6}), the program goes as follows.  Find a
representation for the cross section as a function of energy $\sqrt s$ and
transverse momentum $p_{\perp}$ from experimental data taken at
$\theta^\ast=90^{\circ}$. $F(\sqrt s, p_{\perp})$ is then
completely determined,  because $(\sin\theta^\ast)^D$ is unity at 
$\theta^\ast=90^{\circ}$.

At $\theta^\ast = 90^{\circ}$ the data is well represented by 
\begin{eqnarray}
E\frac{d^3\sigma}{d^3p}(\theta^\star =90^\circ) \equiv F(\sqrt{s},p_{\perp}),
\nonumber 
\end{eqnarray}
with 
\begin{eqnarray}
F(\sqrt{s},p_{\perp}) = \ln(\frac{\sqrt{s}}{\sqrt s_{min}})G(q,p_{\perp}),
\label{F} 
\end{eqnarray}
where $q = s^\frac{1}{4}$ and the COM pion production threshold energy
$\sqrt s_{min} = 2m_p+m_{\pi}$.
The function,
\begin{eqnarray} 
G(q,p_{\perp}) \equiv \frac{E\frac{d^3\sigma}{d^3p}(\theta^\star =90^\circ)}
{\ln(\frac{\sqrt{s}}{\sqrt s_{min}})} \nonumber  
\end{eqnarray}

\noindent 
 was parameterized as
\begin{eqnarray}
G(q,p_{\perp}) &=&\exp\{k_1+k_2p_{\perp}+k_3q^{-1}+k_4p_{\perp}^2+k_5q^{-2} 
+k_6p_{\perp}q^{-1}+k_7p_{\perp}^3+k_8q^{-3} \nonumber \\
& & +k_9p_{\perp}q^{-2}+\ k_{10}p_{\perp}^2q^{-1} +k_{11}p_{\perp}^{-3} \} 
\label{1-8}
\end{eqnarray}
with
$k_1=3.24$, $k_2=-6.046$, $k_3=4.35$, $k_4=0.883$, $k_5=-4.08$,
$k_6=-3.05$, $k_7=-0.0347$, $k_8=3.046$, $k_9=4.098$, $k_{10}=-1.152$, and
$k_{11}=-0.0005$. 
The parameters $k_1 -k_{10}$ were obtained using the numerical curve fitting 
software Table Curve 3D v3 \cite{tc3d} and the eleventh term was added to 
modify the low $p_\perp$ behavior of the parameterization.

With $F(\sqrt s, p_{\perp})$ determined, the function $D(\sqrt s,
p_{\perp},\theta^\ast)$ is the only remaining unknown.  Solving for $D$ yields
\begin{equation}
D(\sqrt s, p_{\perp},\theta^\ast)=\frac{\ln(E\frac{d^3\sigma}{d^3p})-
\ln(F(\sqrt s, p_{\perp}))}{\ln(\sin\theta^\ast)}
\label{1-7}
\end{equation}

Equations (\ref{F}) and (\ref{1-8}) were then used in equation (\ref{1-7}) to 
calculate
values of $D(\sqrt s, p_{\perp}, \theta^\ast)$.  If the function
$D$ is independent of angle, then equation (\ref{1-7}) could be determined
for any fixed angle, $\theta^\ast \neq 90^{\circ}$.  Data were
compared for a range of angular values, and this data revealed that
the function $D$ is not independent of angle.
The angular dependence turned out to be of the form
$(\sin\theta^\ast)^{-0.45}$, and
\begin{equation}
D(\sqrt s, p_{\perp},\theta^\ast)  = (\sin\theta^\ast)^{-0.45} [ c_1 p_{\perp}^{c_2} (\sqrt s)^{c_3} 
+ c_4 \frac{p_{\perp}}{\sqrt s}+\frac{c_5}{\sqrt s}+\frac{1.0}{s}] 
\label{1-9}
\end{equation}
with $c_1=205.7$, $c_2=3.308$, $c_3=-2.875$, $c_4=10.43$, and $c_5=0.8$.

\noindent
{\em The final form of our resultant parameterization for the neutral pion
invariant cross section in proton-proton collisions is equation
(\ref{1-6}) with 
$F(p_{\perp}, \sqrt s)$ given in equation (\ref{F}), $G(q,p_{\perp})$ 
given in equation(\ref{1-8}), and $D(p_{\perp},\sqrt s, \theta^\ast)$ 
given in equation(\ref{1-9}). }
This form is accurate over a much greater range
of transverse momentum values than those covered by previous
representations. Figures  \ref{6.7new} - \ref{53new} show a few 
comparisons. A much more extensive set of data was used in the development 
and comparison of the parameterizations, but they are not shown in this
paper due to space considerations.   
For the low transverse momentum region where the 
cross section is the greatest, the fit is quite similar to that of 
Stephens {\em et al.} \cite{Stephens}.  Also, Figure \ref{tcspi0} shows
that both formulae (\ref{1-5} and \ref{1-6}) integrate to approximately the 
same total cross section, 
which is in agreement with the data from Whitmore {\em et al.} \cite{tcsdat}. 
(Equation \ref{1-6} integrated into a total cross section is denoted as 
Kruger in Figure \ref{tcspi0}.)
A more complete comparison of the integrated total cross section to data is 
given by Stephens {\em et al.}  \cite{Stephens}. Note however that equation
(\ref{1-6}) was based  mainly on the data from  \cite{Fidecaro} - \cite{Stephens}.
Equation (\ref{1-6}) could therefore give unpredictable results
in regions not included in those data sets, particularly for very low
transverse momentum or $\sqrt{s} \gg 63$ GeV.

   \noindent
{\bf 2.2 Charged Pions}

     The available data for charged pions, is less extensive than
$\pi^{\circ}$ data. There is therefore a higher degree of uncertainty in
LIDCS for charged pions.  Integration of a LIDCS to get a total cross
section and comparison of the results to total cross section data, allows a
check of the global fit of a parametrization.  This check was
made for charged as well as neutral pions, but due to a lack of data, it is
more important for charged pions. Parametrizations that do not integrate 
to the correct total cross section can be ruled out, even if the LIDCS data
is well represented, because the global behavior of the parametrization
cannot be accurate. However, producing a correct total cross section upon
integration does not necessarily imply that the global behavior of the
parameterization is correct.
A tighter constraint could be placed on possible LIDCS parametrizations,
if more measurements were made. If the spectral distribution is
measured at three different values of pion energy for two different proton
collision energies, the general behavior of the spectral distribution
could be checked. The angular dependence of LIDCS parametrization
could then be tested by integrating over angle, and comparing the results to
the spectral distribution data. For the purposes of space radiation shielding, 
measurements at proton lab kinetic energies of 3 GeV and
6 GeV, and pion lab kinetic energies of 0.01 GeV, 0.1 GeV, and 1 GeV
would be useful, because this is the region with both a large cross
section, and large Galactic Cosmic Ray fluxes. With these facts in mind, 
a comparison of LIDCS parametrizations with data from
\cite{Busser76,Alper,cdata1,cdata2,Albrow} for charged pion production follows.

     A parametrization for $\pi^-$ of the form
\begin{equation}
E\frac{d^3\sigma}{d^3p}=A\exp(-Bp_\perp^2)
\label{albrow}
\end{equation}

\noindent
has been given by Albrow {\em et al.} \cite{Albrow} where A 
 and B are tabulated functions of $x^\star_R \equiv 
\frac{p^\star}{p^\star_{max}}$. A and B are given only for $x^\star_R=0.18$,  
$x^\star_R=0.21$,
and $x^\star_R=0.25$ which limits the usefulness of this parametrization.

\begin{figure}[htb]
\centering 
\psfig{file=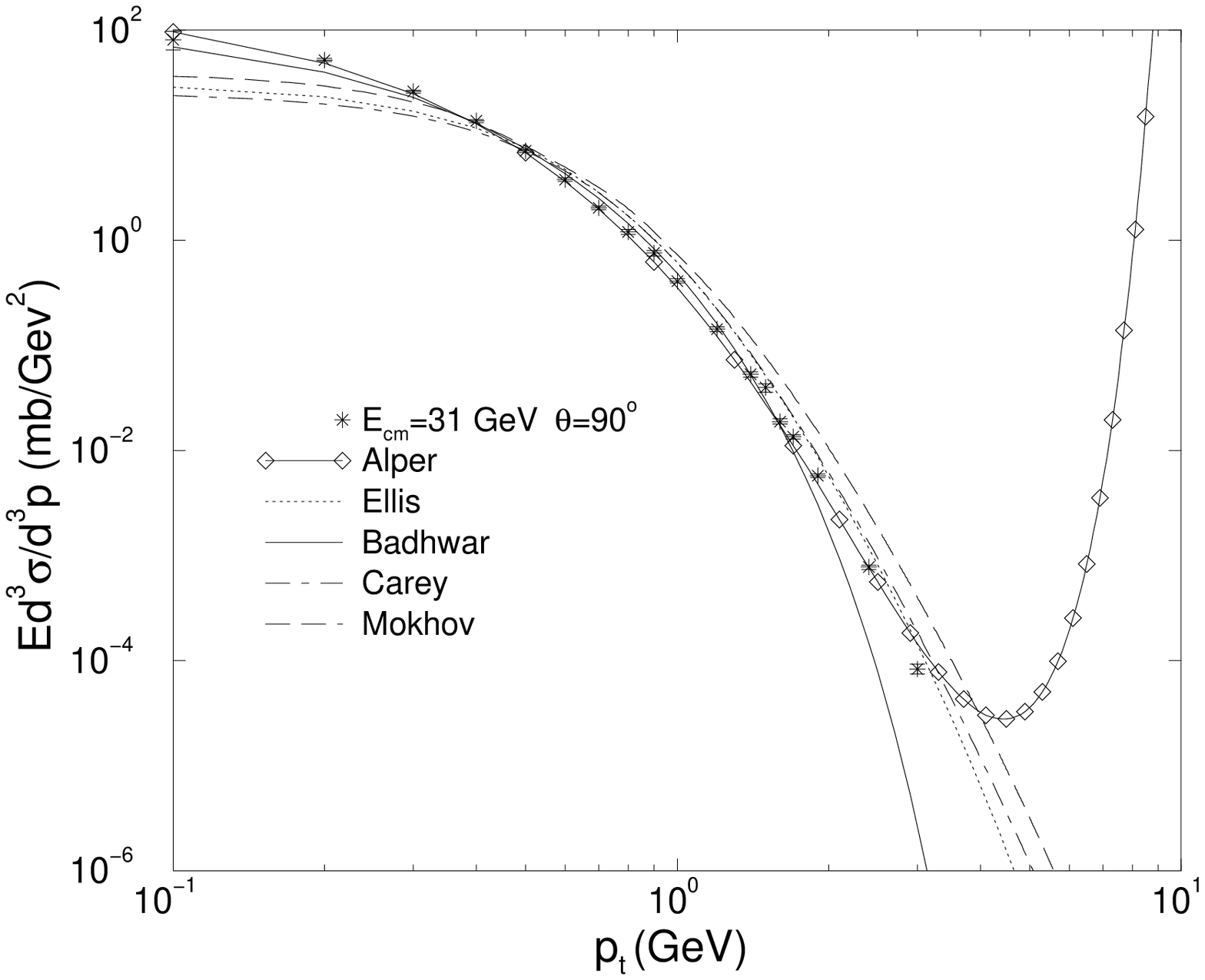,height=3in,width=6in}
%\vspace{.5in}
%\psfig{file=pi-lidcs.dir/lin31.ps,height=3in,width=6in}
\caption{$\pi^-$ production. Same as Figure 7 except $E_{cm}= 31$ GeV, and the parametrization of  Alper {\em et al.} (equation 13) \cite{Alper} is included.} \label{-31}
\end{figure}

\begin{figure}[htb]
\centering 
\psfig{file=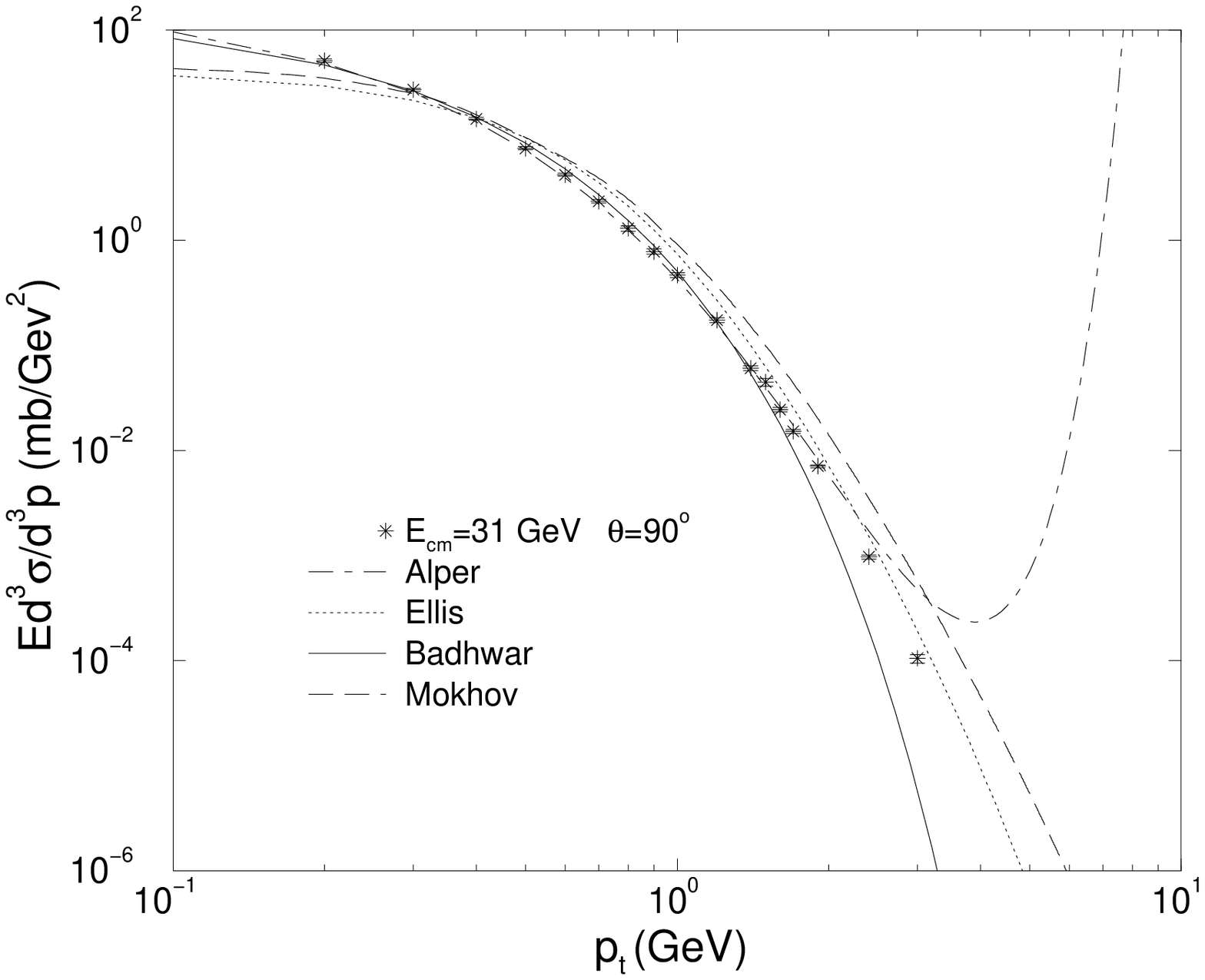,height=3in,width=6in}
%\vspace{.5in}
%\psfig{file=pi+lidcs.dir/lin31.ps,height=3in,width=6in}
\caption{$\pi^+$ production. Same as Figure 9 except  $E_{cm}=31$ GeV, and the parametrization of Alper {\em et al.} (equation 13) \cite{Alper} is included.}  \label{+31}
\end{figure}

Alper {\em et al.} \cite{Alper} have fitted the data for both $\pi^+$ and 
$\pi^-$ production to the following form 
\begin{equation}
E\frac{d^3{\sigma}}{d^3p}=A \exp(-Bp_\perp+Cp_\perp^2) \exp(-Dy^2)
\label{alper}
\end{equation}

\noindent
where $y$ is the longitudinal rapidity, and A, B, C, and D are tabulated
functions of $s$, that are also dependent on the type of produced particle
($\pi^+$ or $\pi^-$). (Note that at $\theta^*=90^\circ$ we have  $y=0$.)
The fit to the
data is excellent for low transverse momentum, as can be seen in Figures
\ref{-31} and \ref{+31},  but these figures also show
that this form has an increasing cross section for high $p_\perp$, which
contradicts the trend in the data. Also, there are different sets of
constants for each different energy, which makes a generalization to
arbitrary energies difficult.
  
\begin{figure}[htb]
\centering 
\psfig{file=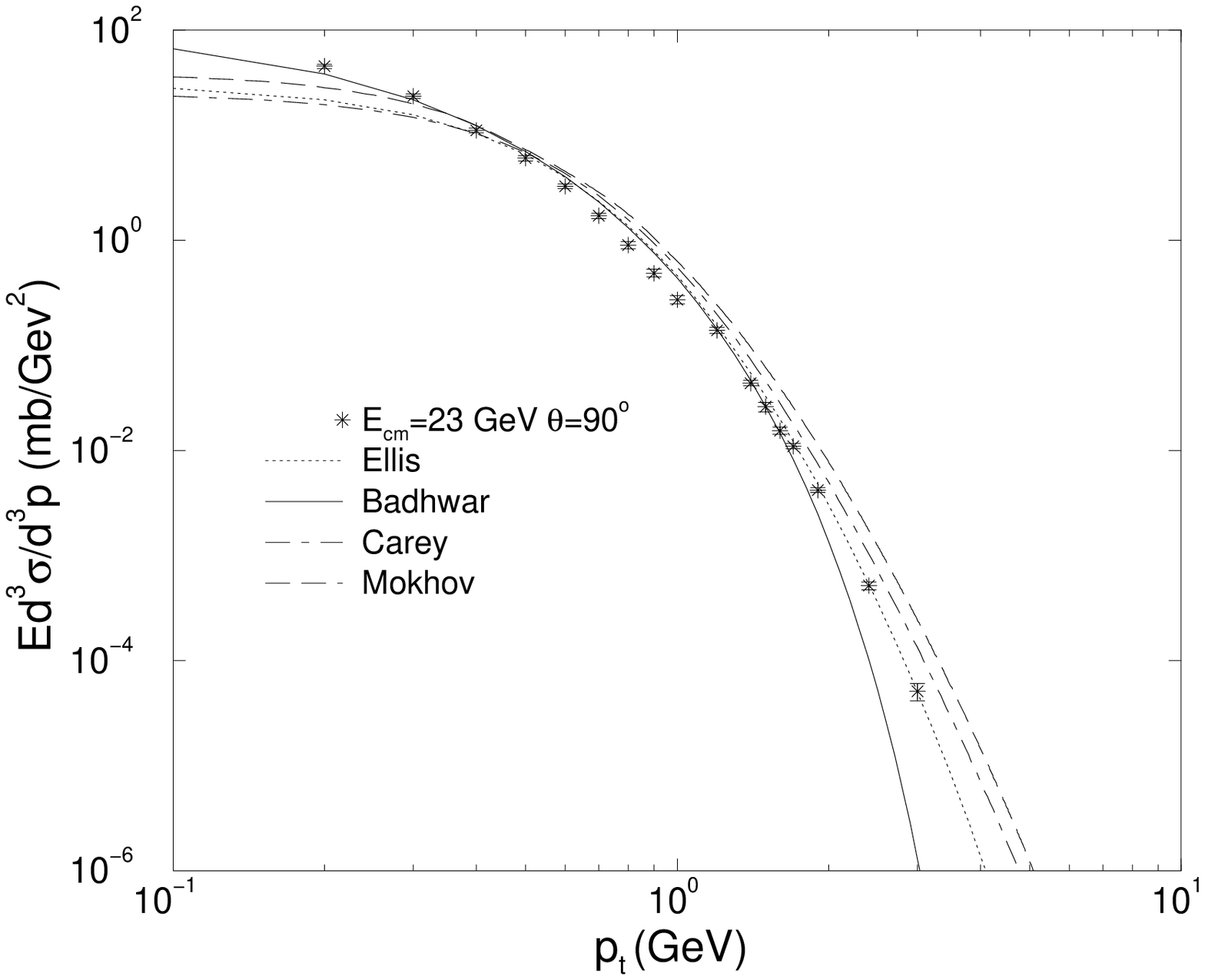,height=3in,width=6in}
%\vspace{.5in}
%\psfig{file=pi-lidcs.dir/lin23.ps,height=3in,width=6in}
\caption{$\pi^-$ production parametrizations of  Ellis {\em et al.} (equation 15) \cite{Ellis77}, Badhwar {\em et al.} (equation 16) \cite{Badhwar}, Carey {\em et al.} (equation 14) \cite{Carey1pi+}, and of Mokhov {\em et al.} (equation 17) \cite{Mokhov} plotted with LIDCS data from \cite{Alper}. LIDCS is plotted against transverse momentum for COM energy $E_{cm}=23$ GeV and pion COM scattering angle $\theta^\ast = 90^{\circ}$} \label{-23}
\end{figure}

\begin{figure}[htb]
\centering 
\psfig{file=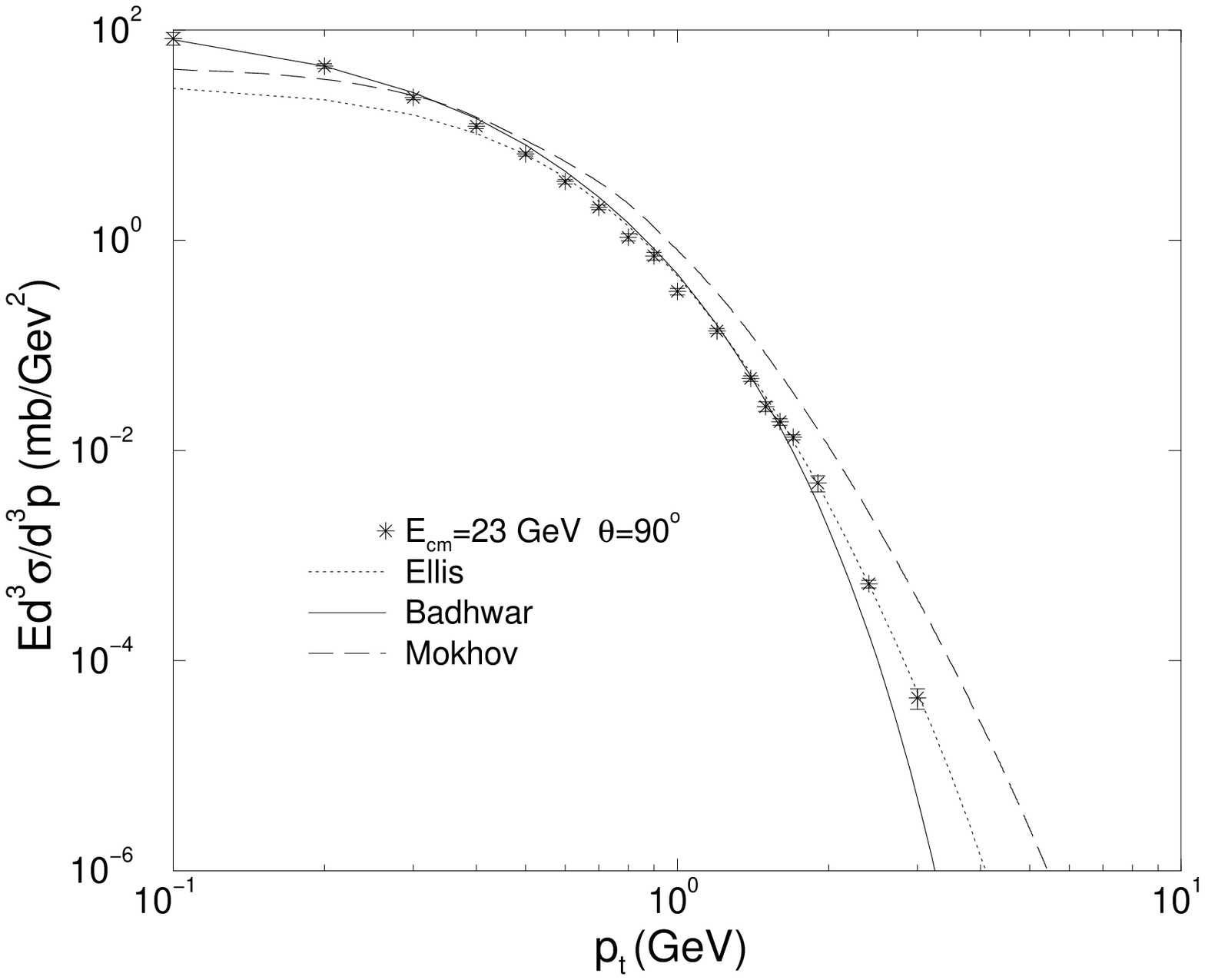,height=3in,width=6in}
%\vspace{.5in}
%\psfig{file=pi+lidcs.dir/lin23.ps,height=3in,width=6in}
\caption{$\pi^+$ production parametrizations of Ellis {\em et al.} (equation 15) \cite{Ellis77}, Badhwar {\em et al.} (equation 16) \cite{Badhwar} and of Mokhov {\em et al.} (equation 17) \cite{Mokhov} plotted with LIDCS data from \cite{Alper}. LIDCS is plotted against transverse momentum for COM energy $E_{cm}=23$ GeV and pion COM scattering angle $\theta^\ast = 90^{\circ}$} \label{+23}
\end{figure}

Parametrizations done by Carey  {\em et al.} \cite{Carey1pi+} and Ellis
{\em et al.} \cite{Ellis77} have a similar form, although Carey's was 
applied only to $\pi^{-}$. Both underestimate LIDCS for low
$p_\perp$, where the cross section is the largest (see Figures
 \ref{-23}-\ref{+31}). 

\noindent
The following is Carey's parametrization
\begin{equation}
E\frac{d^3\sigma}{d^3p}(\pi^-)=N(p_\perp^2+0.86)^{-4.5}(1-x^\star_R)^4
\label{carey}
\end{equation}
where N=13 is the overall normalization constant, and 
$x^\star_R \equiv \frac{p^\star}{p^\star_{max}} \approx 
\frac{2p^\star}{\sqrt{s}}$.

\noindent
The following is Ellis's parametrization which was applied to both $\pi^+$ 
and $\pi^-$ production at $\theta^\star = 90^\circ$.
\begin{equation}
E\frac{d^3\sigma}{d^3p}=A(p_\perp^2+M^2)^{-N/2}(1-x_\perp)^F
\label{ellis}
\end{equation}
where $M, N, F$ are given constants. $A$ is an unspecified overall
normalization for which we used $A=13$, and $x_\perp \equiv \frac{p_\perp}{p^\star_{max}} \approx \frac{2p_\perp}{\sqrt{s}}$.

The most successful LIDCS parametrization
available for charged pion production was found to be the one developed by
Badhwar {\em et al.} \cite{Badhwar}.  
\begin{equation}
E\frac{d^3{\sigma}}{d^3p}= 
\frac{A(1-\tilde{x})^q}{(1+4m_p^2/s)^r}e^{[-Bp_\perp/(1+4m_p^2/s)]}
\label{1-11} 
\end{equation}

\noindent
where $q$ is a function of ${p_\perp}$ and $s$, such that
\begin{eqnarray}
q&=&(C_1+C_2p_\perp+C_3p_\perp^2)/(1+4m_p^2/s)^{1/2} \hspace{.3in} 
\nonumber 
\end{eqnarray}
and
\begin{eqnarray}
\tilde{x}&\approx&[x_{\parallel}^{\star
2}+\frac{4}{s}(p_\perp^2+m_\pi^2)]^\frac{1}{2} \nonumber
\end{eqnarray} 	 

\begin{figure}[htb]
\centering 
\psfig{file=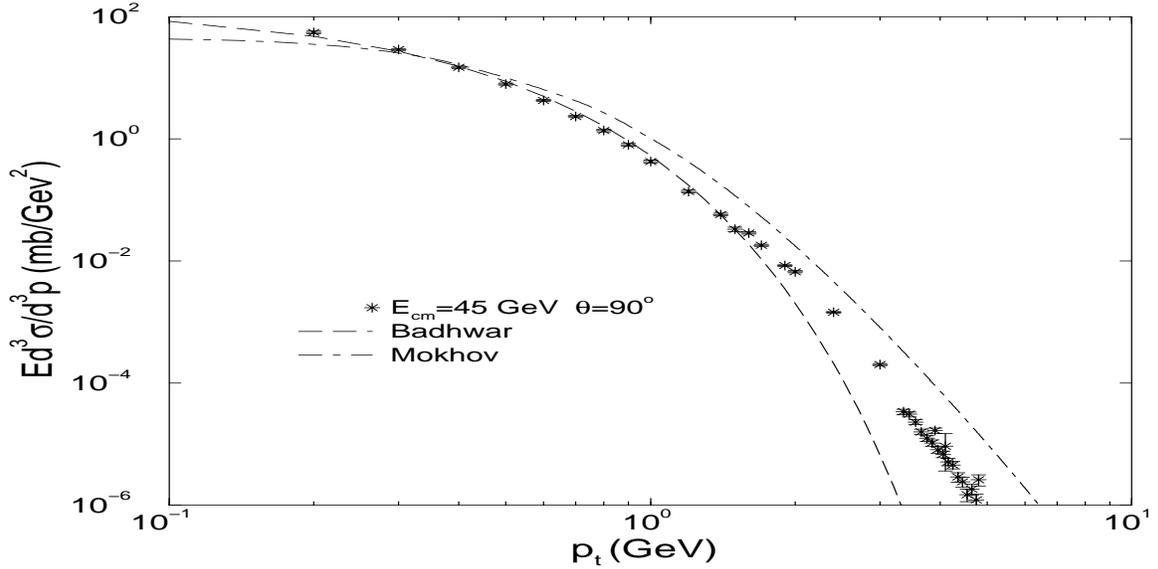,height=3in,width=6in}
%\vspace{.5in}
%\psfig{file=pi+lidcs.dir/lin45.ps,height=3in,width=6in}
\caption{$\pi^+$ production. Same as Figure 9  except  the data is from \cite{Busser76,Alper} for $E_{cm}=$ 45.0 GeV and 44.8 GeV, and the parametrization of Ellis {\em et al.} (equation 15) \cite{Ellis77} is excluded. Parametrizations are plotted at $E_{cm}=45.0$ GeV. }  \label{+45}
\end{figure}

\begin{figure}[htb]
\centering 
\psfig{file=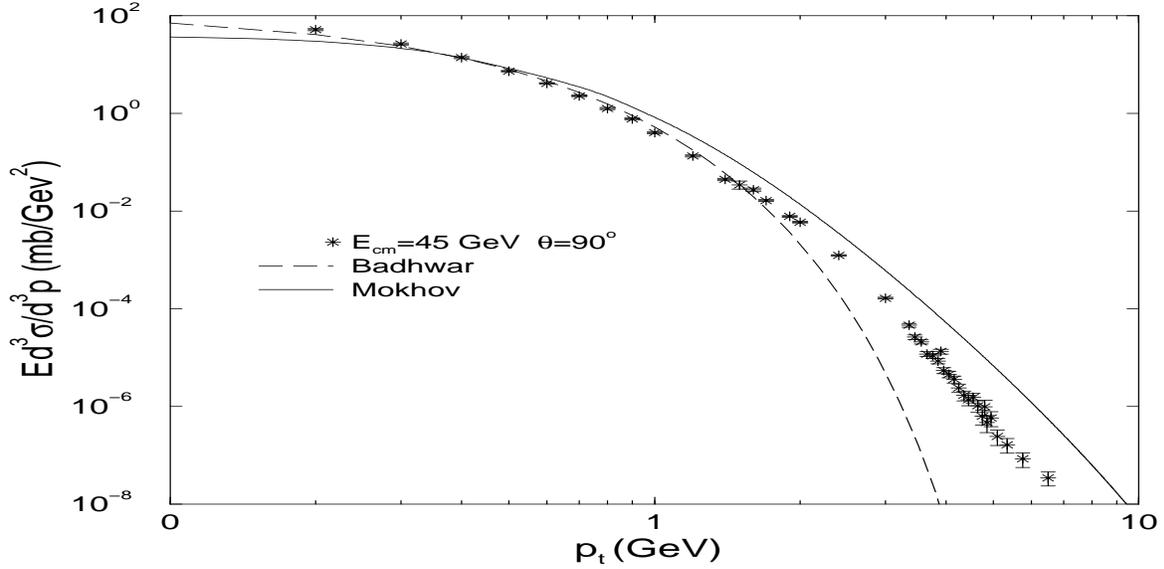,height=3in,width=6in}
%\vspace{.5in}
%\psfig{file=pi-lidcs.dir/lin45.ps,height=3in,width=6in}
\caption{$\pi^-$ production. Same as Figure 7 except the data is from \cite{Busser76,Alper}  for $E_{cm}=$ 45.0 GeV and 44.8 GeV, and some of the parametrizations are excluded. Parametrizations are plotted at $E_{cm}=45.0$ GeV.} 
\label{-45}
\end{figure}

\begin{figure}[htb]
\centering
\psfig{file=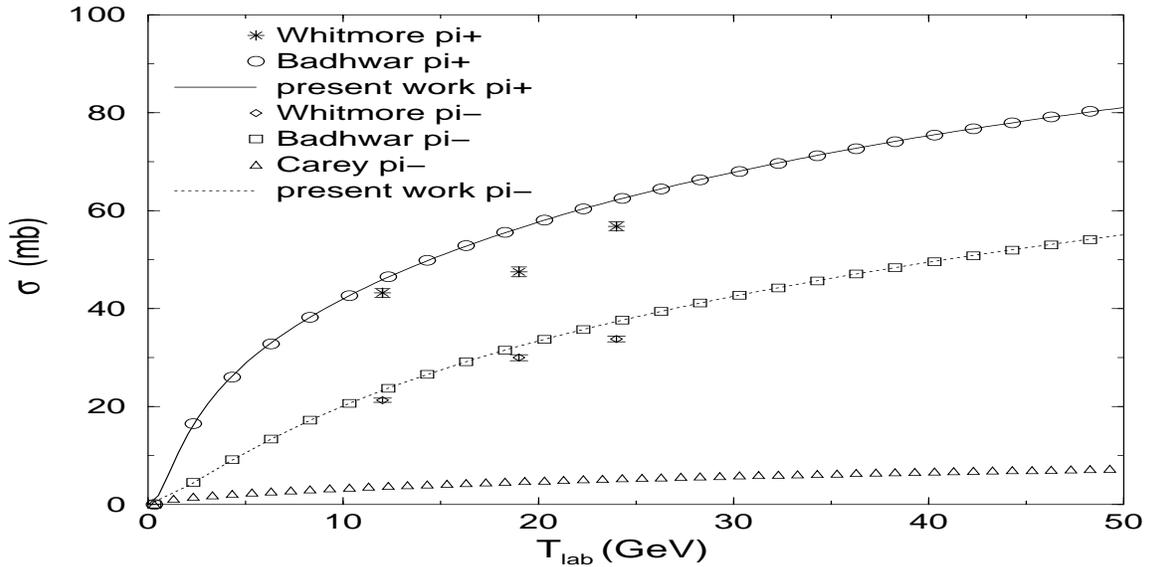,height=3in,width=6in}
\caption{Parametrizations of total $\pi^\pm$ production cross section (present work) (equations 30 and 31) plotted with numerically integrated LIDCS parametrizations of Badhwar {\em et al.} (equation 16) \cite{Badhwar} (circles and squares) and Carey {\em et al.} (equation 14) \cite{Carey1pi+} (triangles).  Six data points are included for comparison (data is from Whitmore 
\cite{tcsdat}).}
\label{tcs}
\end{figure}

\noindent
Here
$x_{\parallel}^{\star}=\frac{p_\parallel^\star}{p^\star_{max}}\approx2
\frac{p_\parallel^{\star}}{\sqrt{s}}$. For $\pi^+$, $A=153 $, $B=5.55 $, 
$C_1=5.3667 $, 
$C_2=-3.5 $, $C_3=0.8334 $, and $r=1$.  For $\pi^-$, $A=127 $, $B=5.3 $, 
$C_1=7.0334 $, 
$C_2=-4.5 $, $C_3=1.667 $, and $r=3 $.  
This form is
accurate for low transverse momentum (Figures \ref{-23}-\ref{-45}), which
is the most important region for radiation shielding due to the
large cross section.  It is also in closed form, so that extra numerical
complexities do not have to be considered. A comparison to a few data
points, shown in Figure \ref{tcs}, demonstrates
that it integrates to the correct total cross section. A more detailed 
comparison of the integrated cross section to experimental data is given by 
Badhwar {\em et al.} \cite{Badhwar}. Because of its
relative accuracy and simplicity, this parametrization was integrated to
get total cross sections and spectral distributions for charged pions.

Mokhov {\em et al.} \cite{Mokhov} have also developed the following formulae
for both $\pi^+$ and $\pi^-$ production.
\begin{equation}
E\frac{d^3\sigma}{d^3p}=A(1-\frac{p^\ast}{p^\ast_{max}})^B 
	\exp(-\frac{p^\ast}{C \sqrt{s}}) V_1(p_\perp) V_2(p_\perp)
\label{mokhov}
\end{equation}
\noindent
where
\begin{eqnarray}
V_1 &=& (1-D) \exp(-E p_\perp^2) + D \exp(-Fp_\perp^2) \hspace{3mm}   \rm{for} 
\hspace{3mm} p_\perp \le 0.933 \, GeV \nonumber \\
    &=&  \frac{0.2625}{(p_\perp^2+0.87)^4} \hspace{3mm}  \rm{for} \hspace{3mm} 
p_\perp >  0.933 \, GeV \nonumber 
\end{eqnarray}
and
\begin{eqnarray}
V_2 &=& 0.7363 \exp(0.875 p_\perp) \hspace{3mm}  \rm{for} \hspace{3mm} p_\perp 
\le 0.35 \, GeV \nonumber \\
    &=& 1 \hspace{3mm}  \rm{for} \hspace{3mm} p_\perp >  0.35 \, GeV \nonumber
\end{eqnarray}
\noindent
with A=60.1, B=1.9, and C=0.18 for $\pi^+$, A=51.2, B=2.6, and C=0.17 for 
$\pi^-$, and D=0.3, E=12, and F=2.7 for both $\pi^+$ and $\pi^-$. 
Figures \ref{-23}-\ref{-45} show that the formula of 
Badhwar has a better fit to the data in the low $p_\perp$ region where 
the cross section is the largest.

\vspace{.15in}
\noindent
{\bf 3. Spectral Distributions and Total Cross Sections}
\vspace{.15in}

\vspace{.15in}
\noindent
{\bf 3.1 Method of Generating Other Cross Sections from a LIDCS}
\vspace{.15in}

While LIDCS contain all the necessary information for a particular process,
sometimes other cross sections are needed. For example, one 
dimensional radiation transport requires probability density distributions
that are integrated over solid angle. These quantities are calculated
in terms of spectral distributions and total cross sections rather than LIDCS,
but with accurate parametrizations of LIDCS, formulae for both spectral 
distributions and total cross sections
can be developed. LIDCS for inclusive pion production
in proton-proton collisions contain dependence on the energy of the
colliding protons ($\sqrt{s}$), on the energy of the produced pion ($T_{\pi}$),
and on the scattering angle of the pion ($\theta$).  Total cross sections 
$\sigma$, which depend only on $\sqrt{s}$, and spectral distributions 
$\frac{d\sigma}{dE}$, which depend on $\sqrt{s}$ and
$T_{\pi}$ can be extracted from a LIDCS by integration.  If azimuthal
symmetry is assumed, these cross sections take the following forms
\begin{equation}
\frac{d\sigma}{dE}  =  2\pi p \int_0^{\theta_{max}} d\theta 
E\frac{d^{3}\sigma}{d^{3}{p}} \sin\theta 
\label{ltosd} 
\end{equation}
\begin{equation}
\sigma  = 2\pi  \int_0^{\theta_{max}} d\theta \int_{p_{min}}^{p_{max}} dp
E\frac{d^{3}\sigma}{d^{3}{p}} \frac{p^{2}\sin\theta}{\sqrt{p^{2}
+ m_{\pi}^2}}
\label{ltotcs}
\end{equation}

\noindent
where $\theta_{max}$, $p_{max}$, and $p_{min}$ are the extrema of the
scattering angle and momentum of the pion respectively, and $m_{\pi}$ is
the rest mass of the pion.  

     In the Center of Mass (COM) frame these extrema can easily be
determined.  Using conservation of momentum and energy, one can easily show
that
\begin{equation}
p^2 = \frac{(s + m_{\pi}^2 - s_x)^2}{4s} - m_{\pi}^2
\end{equation}

\noindent
where $s_x$ is the square of the invariant mass of the sum of all
particles excluding the pion, and $p$ is the magnitude of the three-momentum of
the pion. The independence of $p$ on $\theta$ implies that $\theta$ can
take
on all possible values (ie. $\theta_{max} = \pi$), and the symmetry of the
COM frame implies that $p_{min}$ = 0. For a given value of $s$, it is obvious
that momentum is a maximum when $s_x$ is a minimum. An invariant mass is 
a minimum, when it is equal to the square of the
sum of the rest masses of the particles in question.  Momentum is, therefore,
a maximum when $s_x$ is the square of the sum of the least massive
combination of particles that can be produced while still satisfying all
relevant conservation laws. For the reaction $p + p \rightarrow
\pi + x$, we have   $s_x  \simeq 4m_{p}^2$, where a subscript 
$p$ 
represents a proton.     

If a Lorentz transformation is applied to the maximum COM momentum, the
integration limits can be determined in other frames.  Byckling and Kajantie 
have shown that by transforming to
the lab frame, the following formula can be obtained \cite{Byck}
\begin{equation}
p_{\pi}^{\pm} = [p_a E_{max}^* \sqrt{s} \hspace{.07in} \cos \theta \pm (E_a +
m_{p}) \sqrt{s p_{max}^{*2} - m_{\pi}^2 p_{a}^2 \sin^{2}\theta}] 
	[s + p_{a}^2 \sin^2(\theta)]^{-1}
\end{equation}     

\noindent
where starred quantities are COM variables, and unstarred quantities are
either lab or invariant variables, $m_p$ is the rest mass of a proton; $p_a$ 
is the magnitude of the momentum
of
the projectile proton, and $p^+=p_{max}$ is the maximum pion momentum. The greater
of the two quantities $p^-=p_{min}$ and 0 is the minimum pion momentum, and the
maximum scattering angle can be determined by the requirement that $p^\pm$
be real. This requirement implies that the quantity under the square root
must be greater than or equal to 0.  Solving for $\theta_{max}$ then gives
the formula 
\begin{equation}
\theta_{max} = \sin^{-1}(\frac{\sqrt{s}  \hspace{.07in}
p_{max}^*}{p_a m_{\pi}})
\end{equation}
     With the limits of integration determined, a LIDCS can be turned into
a total cross section or a spectral distribution by numerical integration.
This procedure will, however, give discrete "data" points; not 
closed form expressions. Parametrizations of this numerical data are
needed, if relatively simple formulae for these cross sections are desired.
This process was completed for all three pion species, and the corresponding 
formulae are listed in the next section. It should be noted that the 
accuracy of these parametrizations is limited to that of the original 
LIDCS.

   \vspace{3mm}
\noindent
{\bf 3.2 Parametrizations}
\vspace{3mm}

The surface parametrizations for the spectral distribution as a function
of incident proton kinetic energy in the lab frame ($T_{lab}$) and
the lab kinetic energy of the produced pion ($T_{\pi}$) has been
completed by numerically integrating LIDCS charged pion parametrizations
due to Badhwar {\em et al.} (equation \ref{1-11}) \cite{Badhwar} and the  
neutral pion cross
section both from Stephens and Badhwar (equation \ref{1-5}) \cite{Stephens}, 
and from equation \ref{1-6}. 
The numerical integration routines were checked by computing total 
cross sections in both the lab and COM frames and comparing the results. 
Since total cross section is invariant under the transformation between 
these two frames, the results should be the same in both frames. 
In order to accurately
fit the integration points for low energies, it has been necessary to
consider two regions of the surface and to determine representations for
them individually.  For each of the three pions, the two regions consist of
laboratory kinetic energies ($T_{lab}$) from 0.3 GeV to 2 GeV and from 2 GeV 
to 50 GeV. Using the following parametrizations in energy regions other than 
the region listed above could give unpredictable results since the formulae 
were not tested there.

\newcommand{\X}{T_{lab}}
\newcommand{\Y}{T_\pi}

The neutral pion spectral distribution for the range 0.3 - 2 GeV is
represented by the following equations:  
\begin{eqnarray}
F_2 &=& A_1\Y^{A_2}+A_3\X^{A_4}  \nonumber \\
F_1 &=&
\exp(A_5+\frac{A_6}{\sqrt \X}+A_7\X^{A_8}+A_9\Y^{A_{10}}+
A_{11}\Y^{A_{12}})  \nonumber \\
(\frac{d\sigma}{dE})_{lab} &=& (A_{13}\frac{F_1}{F_2}+A_{14}\exp(A_{16}
\sqrt{\Y}+A_{17}} \Y^{A_{18}}\X^{A_{19}}))\Y^{A_{15}
\label{dsde0a}
\end{eqnarray}
\noindent
 with constants $A_i$ given in Table 1.

{\scriptsize 
\begin{table}[htbp]
\begin{minipage}[t]{3in}
{\scriptsize
  \begin{tabular}{|c|c|c|} \hline 
	$A_1=6.78 \times 10^{-10}$   &   $A_8=-1.75     $  &  $A_{15}=0.25  $ \\ \hline  
        $A_2=-2.86               $   &   $A_9=-32.1     $  &  $A_{16}=-39.4 $ \\ \hline
	$A_3=1.82 \times10^{-8}  $   &   $A_{10}=0.0938 $  &  $A_{17}=2.88  $ \\ \hline
	$A_4=-1.92               $   &   $A_{11}=-23.7  $  &  $A_{18}=0.025 $ \\ \hline
	$A_5=22.3                $   &   $A_{12}=0.0313 $  &  $A_{19}=0.75  $ \\ \hline
	$A_6=0.226               $   &   $A_{13}=2.5 \times 10^{6} $  &             \\ \hline
	$A_7=-0.33               $   &   $A_{14}=1.38   $  &             \\ \hline
  \end{tabular}}
\centering
\parbox{3.0in}{\caption{Constants for equation (23)}} \label{tb1}
\end{minipage}
\hfill
\begin{minipage}[t]{3.0in}
{\scriptsize
  \begin{tabular}{|c|c|c|} \hline
	$B_1=1.3 \times 10^{-10}$   &   $B_8=-1.25      $  &  $B_{15}=60322 $ \\ \hline  
        $B_2=-2.86              $   &   $B_9=-33.2      $  &  $B_{16}=1.07 $ \\ \hline
	$B_3=4.27 \times 10^{-9}$   &   $B_{10}=0.0938  $  &  $B_{17}=-67.5 $ \\ \hline
	$B_4=-2.4               $   &   $B_{11}=-23.6   $  &             \\ \hline
	$B_5=22.3               $   &   $B_{12}=0.0313  $  &             \\ \hline
	$B_6=-1.87              $   &   $B_{13}=2.5 \times 10^{6} $  &             \\ \hline
	$B_7=1.28               $   &   $B_{14}=0.25    $  &
\\ \hline
  \end{tabular}}
\centering 
\parbox{3.0in}{\caption{Constants for equation (24)}} \label{tb2}
\end{minipage}
\end{table}
}
The neutral pion spectral distribution for the range 2 - 50 GeV
is represented by the following equations:
\begin{eqnarray}
F_2 &=& B_1\Y^{B_2}+B_3\X^{B_4}  \nonumber \\
F_1 &=& \exp(B_5 + \frac{B_6}{\sqrt \X}+
       B_7\X^{B_8}+B_9\Y^{B_{10}}+B_{11}\Y^{B_{12}}) \nonumber \\
(\frac{d\sigma}{dE})_{lab}&=& B_{13}\Y^{B_{14}} \frac{F_1}{F_2} +
       B_{15}\Y^{B_{16}}\exp(B_{17} \sqrt \Y)
\label{dsde0b}
\end{eqnarray}
with constants $B_i$ given in Table 2.

The positively charged pion spectral distribution for the range 0.3 - 2 GeV
is represented by the following equations:
\begin{eqnarray}
F_2 &=& C_1\Y^{C_2}+C_3\X^{C_4} \nonumber \\
F_1 &=& \exp(C_5+ \frac{C_6}{\sqrt \X}+C_7\X^{C_8}+C_9\Y^{C_{10}}
      +C_{11}\Y^{C_{12}}\X^{C_{13}}+C_{14} \ln \X) \nonumber \\
(\frac{d\sigma}{dE})_{lab} &=& C_{15}\Y^{C_{16}}\frac{F_1}{F_2}+
C_{17}\Y^{C_{18}}\exp(C_{19}\sqrt{\Y}+C_{20} \sqrt{\X})
\end{eqnarray}
with constants $C_i$ given in Table 3.

{\scriptsize
\begin{table}[htbp]
\begin{minipage}[t]{3.0in}
{\scriptsize
  \begin{tabular}{|c|c|c|} \hline
	$C_1=2.2 \times 10^{-8}$   &  $C_8=-1.75    $  &  $C_{15}=2.5 \times 10^{6}  $ \\ \hline  
        $C_2=-2.7             $   &   $ C_9=-29.4  $  &  $C_{16}=0.25
$ \\ \hline
	$C_3=4.22 \times10^{-7}$   &   $C_{10}=0.0938 $  &  $C_{17}=976                $ \\ \hline
	$C_4=-1.88             $   &   $C_{11}=-24.4   $  &  $ C_{18}=2.3               $ \\ \hline
	$C_5=22.3              $   &   $C_{12}=0.0312 $  &  $ C_{19}=-46                $ \\ \hline
	$C_6=1.98              $   &   $C_{13}=0.0389  $  &  $ C_{20}=-0.989              $ \\ \hline
        $C_7=-0.28             $  &   $C_{14}=1.78   $    &  \\ \hline
\end{tabular}}
\centering
\parbox{3.0in}{\caption{Constants for equation (25)}} \label{tb3}
\end{minipage}
\hfill
\begin{minipage}[t]{3.0in}
{\scriptsize
  \begin{tabular}{|c|c|c|} \hline
	$D_1=4.5 \times 10^{-11}$   &   $D_7=-35.3    $  &  $D_{13}=60322 $ \\ \hline  
        $D_2=-2.98              $   &   $D_8=0.0938   $  &  $D_{14}=1.18   $ \\ \hline
	$D_3=1.18 \times 10^{-9}$   &   $D_9=-22.5    $  &  $D_{15}=-72.2 $ \\ \hline
	$D_4=-2.55              $   &   $D_{10}=0.0313$  &  $D_{16}=0.941 $ \\ \hline
	$D_5=22.3               $   &   $D_{11}=2.5\times 10^6   $  &  $D_{17}=0.1   $ \\ \hline
	$D_6=-0.765             $   &   $D_{12}=0.25  $  &                  \\ \hline
  \end{tabular}}
\centering 
\parbox{3.0in}{\caption{Constants for equation (26)}} \label{tb4}
\end{minipage}
\end{table}
}

The positively charged pion spectral distribution for the range 2 - 50 GeV
is represented by the following equations:
\begin{eqnarray}
F_2 &=& D_1\Y^{D_2}+D_3\X^{D_4} \nonumber \\
F_1 &=& \exp(D_5+ \frac{D_6}{\sqrt \X} +D_7\Y^{D_8}+D_9\Y^{D_{10}})\nonumber \\
(\frac{d\sigma}{dE})_{lab} &=& D_{11}\Y^{D_{12}}\frac{F_1}{F_2}+ 
D_{13}\Y^{D_{14}}\exp(D_{15} \sqrt{\Y} + D_{16} \X^{D_{17}})
\end{eqnarray}
\noindent
with constants $D_i$ given in Table 4.

The negatively charged pion spectral distribution for the range 0.3 - 2 GeV
is represented by the following equations:
\begin{eqnarray}
F_2 &=& G_1\Y^{G_2}+G_3\X^{G_4} \nonumber \\
F_1 &=&\exp(G_5+ \frac{G_6}{\sqrt \X}+G_7\Y^{G_8}+G_9\Y^{G_{10}}) \nonumber \\
(\frac{d\sigma}{dE})_{lab} &=& \Y^{G_{11}}(G_{12}\frac{F_1}{F_2}+
                               G_{13}\exp(G_{14} \sqrt \Y))
\label{dsdeb}
\end{eqnarray}
\noindent
with constants $G_i$ given in Table 5.

{\scriptsize
\begin{table}[htbp]
\begin{minipage}[t]{3.0in}
{\scriptsize
  \begin{tabular}{|c|c|c|} \hline
	$G_1=1.06 \times 10^{-9}$   &   $G_6=-1.5       $  &  $G_{11}= 0.25           $ \\ \hline  
        $G_2=-2.8               $   &   $G_7=-30.5      $  &  $G_{12}= 2.5 \times 10^6$ \\ \hline
	$G_3=3.7 \times10^{-8}  $   &   $G_8=0.0938     $  &  $G_{13}=7.96            $ \\ \hline
	$G_4=-1.89              $   &   $G_9=-24.6      $  &  $G_{14}=-49.5           $ \\ \hline
	$G_5=22.3               $   &   $G_{10}=0.0313  $  &                            \\ \hline
  \end{tabular}}
\centering
\parbox{3.0in}{\caption{Constants for equation (27)}} \label{tb5}
\end{minipage}
\hfill
\begin{minipage}[t]{3.0in}
{\scriptsize
  \begin{tabular}{|c|c|c|} \hline
	$H_1=2.39 \times 10^{-10}$   &   $H_7=-31.3               $  &  $H_{13}=60322 $ \\ \hline  
        $H_2=-2.8             $   &   $H_8=0.0938               $  &  $H_{14}=1.1   $ \\ \hline
	$H_3=1.14 \times 10^{-8}$   &   $H_9=-24.9                $  &  $H_{15}=-65.9 $ \\ \hline
	$H_4=-2.3               $   &   $H_{10}=0.0313            $  &  $H_{16}=-9.39 $ \\ \hline
	$H_5=22.3               $   &   $H_{11}=2.5 \times 10^{6} $  &  $H_{17}=-1.25 $ \\ \hline
	$H_6=-2.23              $   &   $H_{12}=0.25              $  &                  \\ \hline
\end{tabular}}
\centering 
\parbox{3.0in}{\caption{Constants for equation (28)}} \label{tb6}
\end{minipage}
\end{table}
}

The negatively charged pion spectral distribution for the range 2 - 50 GeV
is represented by the following equations:
\begin{eqnarray}
F_2 &=& H_1\Y^{H_2}+H_3\X^{H_4} \nonumber \\
F_1 &=&\exp(H_5+  \frac{H_6}{\sqrt \X} +H_7\Y^{H_8}+H_9\Y^{H_{10}}) \nonumber \\
(\frac{d\sigma}{dE})_{lab} &=& H_{11}\Y^{H_{12}}\frac{F_1}{F_2}+
                            H_{13}\Y^{H_{14}}\exp(H_{15} \sqrt{\Y} + 
	H_{16}\X^{H_{17}})
\end{eqnarray}
\noindent
with constants $H_i$ given in Table 6.

 Total inclusive cross sections are represented by the following equations.

\begin{equation}
% old para \sigma_{\pi^0}= -6.25-0.000398\X^{2.5}+11.3\X^{0.5}
\sigma_{\pi^0}=(0.007 + 0.1 \frac{\ln(\X)}{\X} + \frac{0.3}{\X^2})^{-1}
\label{tp0}
\end{equation}
\begin{equation}
\sigma_{\pi^+} = (0.00717+0.0652\frac{\ln(\X)}{\X}+\frac{0.162}{\X^2})^{-1}
\label{tp+}
\end{equation}
\begin{equation}
\sigma_{\pi^-} = (0.00456+\frac{0.0846}{\X^{0.5}}+\frac{0.577}{\X^{1.5}})^{-1}
\label{tp-}
\end{equation}

\vspace{3mm}
{\em For neutral pions, spectral distributions and total cross sections that 
were based on our own parametrization given in equation (\ref{1-6}) were 
also developed}. The formula for the 
spectral distribution was not divided into two regions, and it is much simpler 
than the previous formulae.
%
%old para
%\begin{equation}
%(\frac{d\sigma}{dE})_{lab}=\exp(K_1 + K_2 \ln(\X^{0.6}) +K_3\X^{-1.2} + 
%K_4\Y^{0.2} + K_5\Y^{0.1} \ln(\Y^{0.2}))
%\label{dsdef}
%\end{equation}
%
\begin{equation}
(\frac{d\sigma}{dE})_{lab}=\exp(K_1 +\frac{K_2}{\X^{0.4}} + 
\frac{K_3}{\Y^{0.2}} + \frac{K_4}{\Y^{0.4}})
\label{dsdef}
\end{equation}
\noindent
where $K_1 = -5.8$, $K_2 = -1.82$, $K_3 = 13.5$, $K_4 = -4.5$.

Because equation (\ref{1-6}) and Stephens LIDCS parametrization integrate 
to nearly the same total cross section (see Figure \ref{tcspi0}), separate 
total cross section parametrizations are not necessary (i.e. use equation 
\ref{tp0}).

\vspace{3mm}
\noindent
{\bf 3.3 Discussion of Figures}
\vspace{3mm}

As discussed previously, Figures \ref{6.7new} - \ref{53new}
show a comparison of LIDCS parametrizations for
$\pi^\circ$ production of  Carey {\em et al.} (equation \ref{1-4}) 
\cite{Carey76}, Stephens 
{\em et al.} (equation \ref{1-5}) \cite{Stephens}, and equation (\ref{1-6}) 
plotted with data from 
\cite{Fidecaro} - \cite{Stephens}. The figures are graphs of cross section plotted against 
transverse momentum ($p_\perp$) for various values of COM energy ($E_{cm}$)
and COM scattering angle ($\theta^\ast$).  
Figure \ref{6.7new} shows that the
parametrization of Carey {\em et al.} is not an adequate representation
of the data. Figures \ref{62.6new} and \ref{53new} show that the
parametrization of Stephens {\em et al.} fails for high transverse
momentum by severely underpredicting the cross section. 

Figure \ref{tcspi0} shows  numerically integrated LIDCS parametrizations 
of Stephens
et al. (equation \ref{1-5}) \cite{Stephens}, of Carey et al. (equation 
\ref{1-4}) \cite{Carey76}, and of equation (\ref{1-6}) (referred to as
Kruger) for $\pi^0$ production plotted with a parametrization of the
integrated formulae of  Stephens et al.  referred to as Stephens-total-param
(equation \ref{tp0}). Three data points from Whitmore \cite{tcsdat}
show that Carey's parametrization does not integrate to the correct
values and that the rest are quite accurate (see \cite{Stephens} for more 
detail). 

\begin{figure}[htb]
\centering
%\psfig{file=graphs.dir/dsdepi0lin.eps,height=2.6in,width=6in}
%\vspace{.5in}
\psfig{file=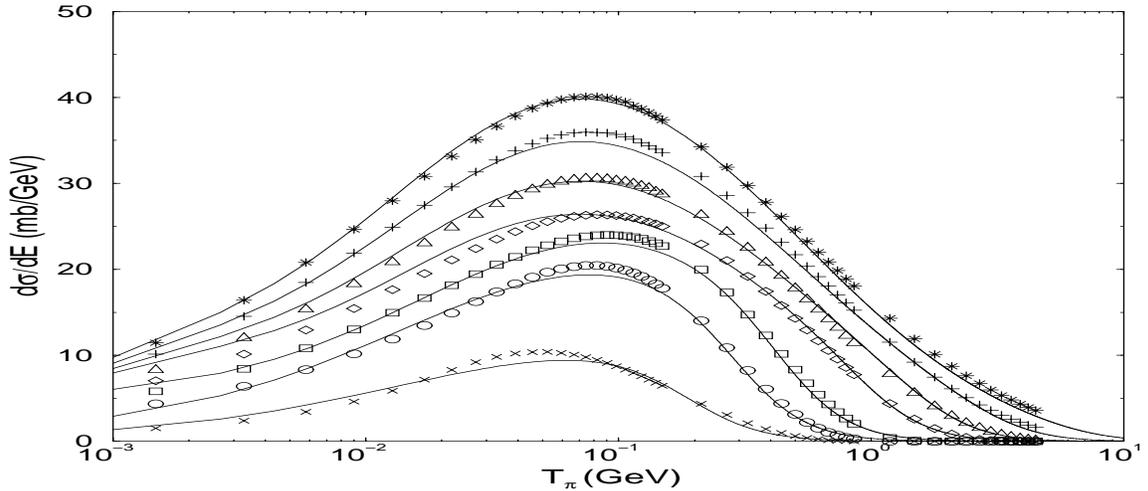,height=2.6in,width=6in}
\caption{$\pi^0$ spectral distribution parametrizations  of equations (23) and (24) (solid lines) plotted with LIDCS parametrization of Stephens (equation 6) \cite{Stephens} numerically integrated at lab kinetic energies of 0.5 GeV, 1.0 GeV, 1.9 GeV, 5.0 GeV, 9.5 GeV, 20 GeV, and 50 GeV, listed in order of increasing cross section (symbols).} \label{dsdepi0}
\end{figure}

\begin{figure}[htb]
\centering
%\psfig{file=graphs.dir/dsdefaustlin.eps,height=2.6in,width=6in}
%\vspace{.5in}
\psfig{file=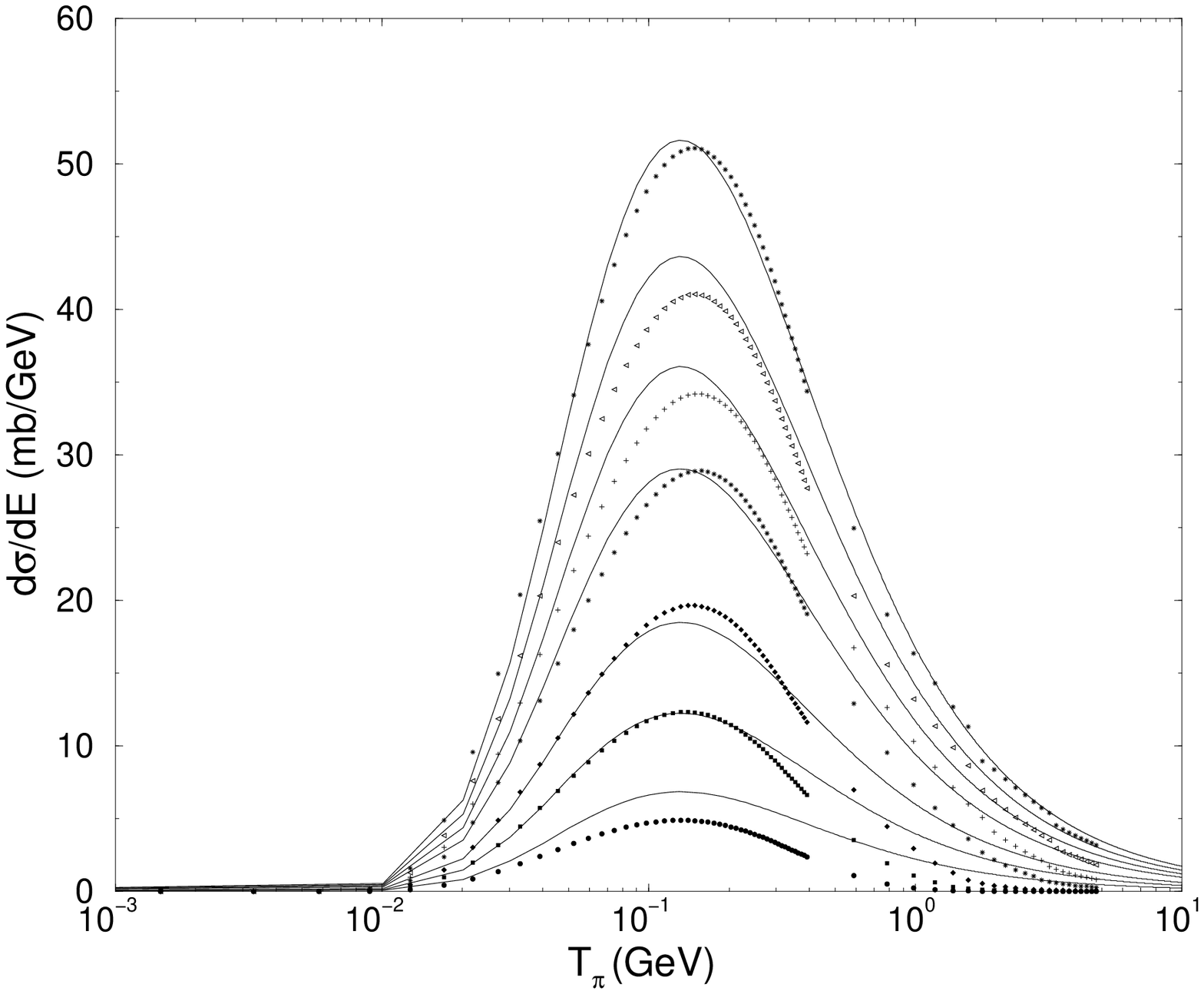,height=2.6in,width=6in}
\caption{$\pi^\circ$ spectral distribution parametrization of equation (32) (solid lines) plotted with equation(7) numerically integrated at lab kinetic energies of 0.5 GeV, 1.0 GeV, 1.9 GeV, 5.0 GeV, 9.5 GeV, 20 GeV, and 50 GeV, listed in order of increasing cross section (symbols). } \label{dsdepi0f}
\end{figure}

Figure \ref{dsdepi0} shows $\pi^0$ spectral distribution parametrizations 
given by equations (\ref{dsde0a}) and (\ref{dsde0b}) plotted with LIDCS
 parametrization
of Stephens (equation \ref{1-5}) numerically integrated at several lab kinetic energies.
Figure \ref{dsdepi0f}  shows $\pi^\circ$
spectral distribution parametrizations given by 
equation (\ref{dsdef})  plotted with the numerical integration of  
equation (\ref{1-6}). The shapes of the two spectral distributions
look quite different even though both original LIDCS formula have a
similar fit to the data at low $p_\perp$ where the cross section is the 
greatest, and both integrate to the same total cross section. This
implies that the available data is not sufficient to tightly constrain 
the shape of the spectral distribution. 

As discussed previously, Figures \ref{-23} - \ref{-45} show $\pi^+$ and 
$\pi^-$ LIDCS  parametrizations of Alper
{\em et al.}  (equation \ref{alper}) \cite{Alper},  Badhwar {\em et al.} 
(equation \ref{1-11}) \cite{Badhwar},
Ellis {\em et al.}  (equation \ref{ellis}) \cite{Ellis77},  Carey {\em et 
al.} (equation \ref{carey})  \cite{Carey1pi+}, and  Mokhov {\em et al.}
 (equation \ref{mokhov}) \cite{Mokhov} and LIDCS
 data from \cite{Busser76, Alper} plotted against transverse momentum 
($p_t \equiv p_\perp$) for different values of COM energy ($E_{cm}$), but all 
at $\theta^\ast=90^\circ$. These graphs show  that the parametrizations of
Badhwar best fit the data, but underpredict the cross section for large 
transverse momentum. 

Figure \ref{tcs} shows the numerically integrated LIDCS 
parametrizations of
Badhwar {\em et al.} (equation \ref{1-11})  \cite{Badhwar}, and of Carey {\em 
et al.} (equation \ref{carey})  \cite{Carey1pi+} for $\pi^+$
and $\pi^-$ plotted with parametrizations of the integrated formulae of
Badhwar referred to as present work  (equations \ref{tp+} and \ref{tp-}).
Three data points from Whitmore {\em et al.} \cite{tcsdat} show that Carey's
parametrization does not integrate to the correct values and that
Badhwar's formula is accurate. The figures also show that the
parametrization fits the numerically integrated formulae very well.
 
\begin{figure}[htb]
\centering 
%\psfig{file=graphs.dir/dsdepi-lin.eps,height=2.1in,width=6in}
%\vspace{.5in}
\psfig{file=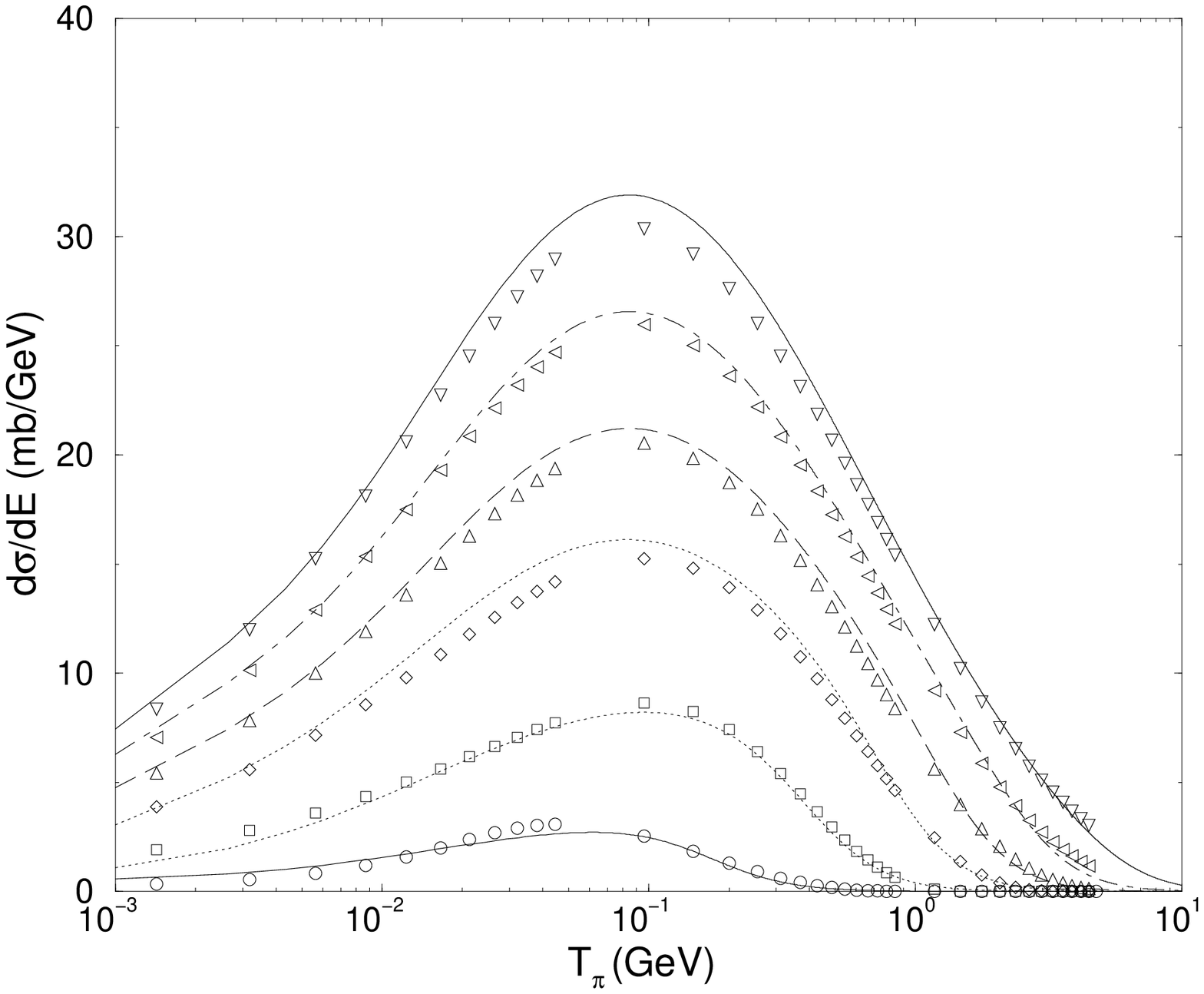,height=2.1in,width=6in}
\caption{$\pi^-$ spectral distribution parametrizations (equations 27 and 28) (solid lines) plotted with LIDCS parametrization of Badhwar {\em et al.} (equation 16) \cite{Badhwar} (symbols) numerically integrated at lab kinetic energies of 0.5 GeV, 1.9 GeV, 5.0 GeV, 9.5 GeV, 20 GeV, and 50 GeV, listed in order of increasing cross section.} \label{dsdepi-}
\end{figure}

\begin{figure}[htb]
\centering
%\psfig{file=graphs.dir/dsdepi+lin.eps,height=2.1in,width=6in}
%\vspace{.5in}
\psfig{file=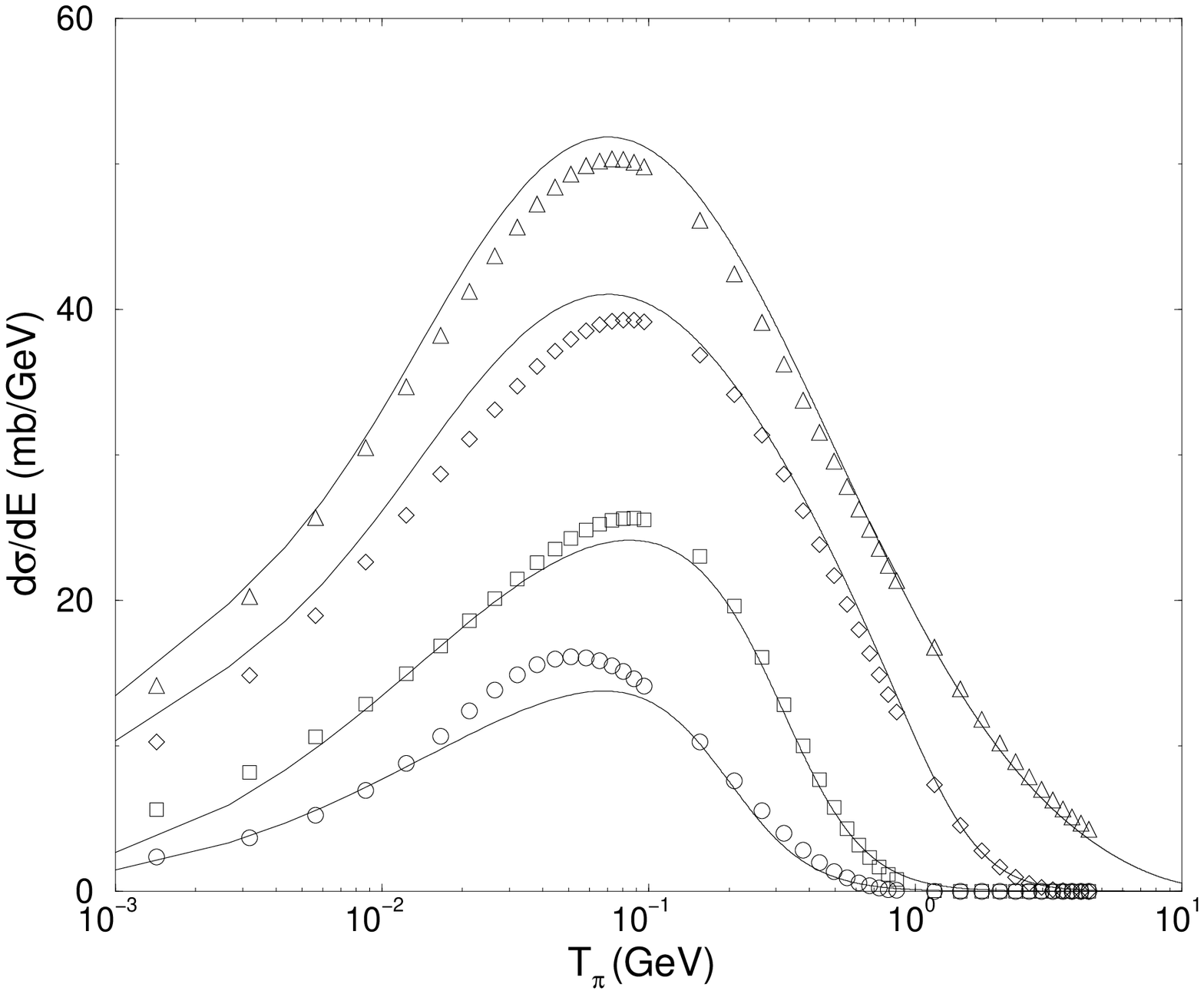,height=2.1in,width=6in}
\caption{$\pi^+$ spectral distribution parametrizations (equations 25 and 26) (solid lines) plotted with LIDCS parametrization of Badhwar {\em et al.} (equation 16) \cite{Badhwar} (symbols) numerically integrated at lab kinetic energies of 0.5 GeV, 1.1 GeV, 5.0 GeV, and 50 GeV, listed in order of increasing cross section. }
\label{dsdepi+}
\end{figure}

Figures \ref{dsdepi-} and \ref{dsdepi+} show $\pi^-$ and $\pi^+$ spectral 
distribution parametrizations plotted with LIDCS 
parametrization of
Badhwar {\em et al.} (equation \ref{1-11})  \cite{Badhwar} numerically 
integrated. The plot is
of cross section ($\frac{d\sigma}{dE}$)  plotted against the
kinetic energy of the produced pion $T_\pi$ at
several values for the lab kinetic energies of the colliding proton. 
The graphs clearly show that the spectral distribution parametrizations 
have excellent fits to the integrated LIDCS parametrizations.

   \vspace{3mm}
\noindent
{\bf 4. Summary and Conclusions}
\vspace{3mm}

This paper presents parametrizations of cross sections for inclusive pion
production in proton-proton collisions. The cross sections of interest are
Lorentz Invariant Differential Cross Sections (LIDCS), lab frame spectral
distributions, and total cross sections. 
For neutral pions the parametrization of Stephens {\em et. al.}
\cite{Stephens} (equation \ref{1-5}) fit the data well for low values of
$p_\perp$, but overpredicted the cross section by many orders of magnitude
at high $p_\perp$ values. Because of this inaccuracy, equation
(\ref{1-6}) was developed.{\em The final form of our resultant 
parametrization for the neutral pion
invariant cross section in proton-proton collisions is equation
(\ref{1-6}) with $D(p_{\perp},\sqrt s, \theta^\ast)$ given in equation
(\ref{1-9}) ,
$F(p_{\perp}, \sqrt s)$ given in equation (\ref{F}), and $G(q,p_{\perp})$ 
given in equation(\ref{1-8}). } This formula is as accurate as that of Stephens {\em et.
al.} \cite{Stephens} at low $p_\perp$ values, but is much more accurate at
high $p_\perp$ values. For charged pions the formulae of Badhwar {\em et al.}
 (equation \ref{1-11}) \cite{Badhwar} were found to best represent the data except at high $p_\perp$ values.
These formulae were used in the development of spectral distributions and total cross sections
because they are the most accurate at low $p_\perp$ where the cross section is 
the greatest.

The data for lab frame spectral distributions and total cross sections is
scarce, so parametrizations for these quantities were developed using the 
above LIDCS
formulae. These formulae were numerically integrated, resulting in
discrete numerical "data" points for these other cross sections.
 The accuracy of the
representations of lab frame spectral distributions and total cross
sections is, therefore, limited to the accuracy of the original LIDCS. The
numerical  "data" was then parametrized so that closed form expressions 
(equations \ref{dsde0a}-\ref{dsdef}) could be obtained.  As a check on the 
accuracy,
the total cross section numerical "data" was compared to experimental
data. They were found to agree quite well, but when the numerical "data"
for the spectral distributions for the formulae for $\pi^0$ production
(equations \ref{dsde0a}-\ref{dsde0b} and \ref{dsdef}) are compared (ie.
compare Figure \ref{dsdepi0} to Figure \ref{dsdepi0f}), 
they are found to disagree. Since both original LIDCS formulae 
fit the data
well at low $p_\perp$ where the cross section is greatest, and both formulae
integrate to the correct total cross section, the available data must not
be sufficient to uniquely determine the global behavior of the LIDCS.
The data for charged pion production is much more limited than the data
for neutral pion production, so the same problem exists for
charged pions. 

To more accurately determine the cross sections for space
radiation applications, measurements of the spectral  distribution at
lower energies (for example, proton lab kinetic energies 3 and 8 GeV, and pion lab
kinetic energies of 0.01 GeV, 0.1 GeV, and 1 GeV) would need to be taken. 
These measurements would put a much tighter constraint
on the global properties of the LIDCS, and the spectral distribution
parametrizations could also be made more accurate.

\vspace{3mm}
\noindent
{\bf Acknowledgements:}
\vspace{3mm}

\noindent 
The authors would like to thank Sean Ahern, and Alfred Tang for their 
help on the project. SRB was supported by the Wisconsin Space
Grant Consortium, NASA grant NCC-1-260, and NASA Graduate Student Researchers
Program Fellowship NGT-52217. ATK and MN
were supported by the Wisconsin Space Grant Consortium, NASA grant 
NCC-1-260, and NSF grant PHY-9507740. JWN and SRS were supported by NASA 
grants NCC-1-260, and NCC-1-354.

   \vspace{.15in}
\noindent
{\bf Appendix A. Synopsis of Data Transformations}
\vspace{.15in}

The data that was used in the comparison of different parametrizations was
given in terms of several different kinematic variables. Some of the LIDCS 
data was 
transformed so that all data would be expressed in terms of the same 
variables: $p_\perp$, $\theta^\ast$, and $E_{cm}=\sqrt{s}$. 
The following is a synopsis of the transformations that were 
performed for the data plotted in the figures.

The data from Carey {\em et al.} \cite{Carey76} was listed for different 
values of $P_p$, $p_\perp$, and $\theta$. $\sqrt{s}$, $p_\perp$, and 
$\theta$ were used by Eggert {\em et al.} 
Stephens {\em et al.} \cite{Stephens} used photon production data from 
Fidecaro {\em et al.} \cite{Fidecaro} to derive pion production cross sections.
The variables $T_{lab}$, $\theta$, and $p$ were used by Stephens.  
 $\sqrt{s}$, $p_\perp$, and the longitudinal rapidity $y$ were
used by Alper {\em et al.} \cite{Alper}, but only data with $y=0$ was used in
the figures. When $y=0$ then $\theta^\ast=90^\circ$. 

The necessary transformations are as follows.
\begin{equation}
p_\perp = p \sin \theta
\end{equation}
$T_{lab}$, $P_p$, and  $\theta$ can be transformed into $\sqrt{s}$
and $\theta^\ast$ by using the following Lorentz transformations to change to 
the COM frame. First express  $T_{lab}$ and $P_p$ as total lab energy $E$.
\begin{eqnarray}
E &=& T_{lab} + m_p
  = \sqrt{P_p^2 + m_p^2}
\end{eqnarray}
Now perform the following Lorentz transformations.
\begin{eqnarray}
E_{cm}  &=& -\gamma v  p \cos \theta + \gamma E \\
\theta^* &=& \tan^{-1}(\frac{p \sin \theta}{\gamma p \cos \theta - 
\gamma v E})
\end{eqnarray}
where
\begin{eqnarray}
\gamma = \frac{T_{lab}+2m_p}{\sqrt{s}} \\
v=\sqrt{1-\gamma^{-2}}
\end{eqnarray}

   %LIDCS PI0
%%%%%%%%%%%%%%%%%%%%%%%%%%%%%%%%%%%%%%%%%%%%%%%%%%%%%%%%%%%%%%%%%%%%%%%%%%%%

%TCS PI0 bad

%DSDE PI0 bad

%DSDE faust

   %% LIDCS 

%%%%%%%%%%%%%%%%%%%%%%%%%%%%%%%%%%%%%%%%%%%%%%%%%%%%%%%%%%%%%%%%%%%%%%%%%%%%%
%TCS NEW

%TCS PI+ bad
%\begin{figure}[htb]
%\centering
%\psfig{file=graphs.dir/tcspi+.eps,height=2.6in,width=6in}
%\caption{Parametrizations of total $\pi^+$ production cross section (present work) (equation 30) plotted with numerically integrated LIDCS parametrizations of Badhwar {\em et al.} (equation 16)  [30] (circles).  Three data points are included for comparison (data is from Whitmore [24]).}
%\label{tcspi+}
%\end{figure}

%TCS PI- bad
%\begin{figure}[htb]
%\centering
%\psfig{file=graphs.dir/tcspi-.eps,height=3in,width=6in}
%\caption{Parametrizations of total $\pi^-$ production cross section (present work) (equation 31) plotted with numerically integrated LIDCS, parametrizations of Badhwar {\em et al.} (equation 16) [30] and of Carey{\em et al.} (equation 14) [29].  Three data points are included for comparison (data is from Whitmore [24]).}
%\label{tcspi-}
%\end{figure}

%
%DSDE PI-

%DSDE PI+ 

\end{document}